\documentclass[aps,prfluids,preprint,amsmath,amssymb,onecolumn,author-numerical,floatfix]{revtex4-2}
\usepackage{graphicx}
\usepackage{newtxtext}
\usepackage{newtxmath}
\usepackage{natbib}
\usepackage{hyperref}
\usepackage{siunitx}
\usepackage{comment}
\hypersetup{
    colorlinks = true,
    urlcolor   = blue,
    citecolor  = black,
}

\newcommand{\RomanNumeralCaps}[1]
\linenumbers

\draft % marks overfull lines with a black rule on the right

\begin{document}

%\preprint{}

\title{Effects of elastoviscoplastic properties of mucus on airway closure in healthy and pathological conditions} %Title of paper



\author{O. Erken}
%\thanks{}
\altaffiliation[Current address: ]{School of Engineering, Institute for Infrastructure and Environment, The University of Edinburgh, United Kingdom}
\affiliation{Department of Mechanical Engineering, Koc University, Istanbul, Turkey}

\author{B. Fazla}
%\thanks{}
\affiliation{Department of Mechanical Engineering, Koc University, Istanbul, Turkey}

\author{D. Izbassarov}
%\thanks{}
\affiliation{Finnish Meteorological Institute, Erik Palmenin aukio 1, 00560 Helsinki, Finland}

\author{F. Roman\`{o}}
%\thanks{}
\affiliation{Univ. Lille, CNRS, ONERA, Arts et M\'etiers Institute of Technology, Centrale Lille, UMR 9014 -LMFL - Laboratoire de M\'ecanique des Fluides de Lille - Kamp\'e de F\'eriet, F-59000 Lille, France}

\author{J. B. Grotberg}
%\email[]{}
%\thanks{}
\affiliation{Department of Biomedical Engineering, University of Michigan, Ann Arbor, MI, 48109, USA}

\author{M. Muradoglu}
\email[Email address for correspondence: ]{mmuradoglu@ku.edu.tr}
%\thanks{}
\affiliation{Department of Mechanical Engineering, Koc University, Istanbul, Turkey}


\date{\today}

\begin{abstract}
Airway mucus is a complex material with both viscoelastic and viscoplastic properties that vary with healthy and pathological conditions of the lung. In this study, the effects of these conditions on airway closure are examined in a model problem, where an elastoviscoplastic (EVP) single liquid layer lines the inner wall of a rigid pipe and surrounds the air core. The EVP liquid layer is modelled using the Saramito-HB model. The parameters for the model are obtained for the mucus in healthy, asthma, chronic obstructive pulmonary disease (COPD) and cystic fibrosis (CF) conditions by fitting the rheological model to the experimental data. Then, the liquid plug formation is studied by varying the Laplace number and undisturbed liquid film thickness.

Airway closure is a surface-tension-driven phenomenon that occurs when the ratio of the pulmonary liquid layer thickness to the airway radius exceeds a certain threshold. In previous studies, it has been found that airway epithelial cells can be lethally or sub-lethally damaged due to the high peak of the wall stresses and stress gradients during the liquid plug formation. Here, we demonstrate that these stresses are also related to the EVP features of the liquid layer. Yielded zones of the liquid layer are investigated for the different mucus conditions, and it is found that the liquid layer is in a chiefly unyielded state before the closure, which indicates that this phase is dominated by the elastic behaviour and solvent viscosity. This is further confirmed by showing that the elastic coefficient is one of the most critical parameters determining whether the closure occurs or not. This parameter also largely affects the closure time. The wall stresses are also investigated for the pathological and healthy cases. Their peaks for COPD and CF are found to be the highest due to the viscoelastic extra stress contribution. Contrary to the Newtonian case, the wall stresses for COPD and CF do not smoothly relax after closure, as they rather remain effectively almost as high as the Newtonian peak. Moreover, the local normal wall stress gradients are smaller for the COPD and CF liquid layer due to their higher stiffness causing a smaller curvature at the capillary wave. The local tangential wall stress gradients are also shown to be smaller for these cases because of the slower accumulation of the liquid at the bulge.
\end{abstract}


\maketitle %\maketitle must follow title, authors, abstract and \pacs

\section{Introduction}\label{sec:introduction}
The inner surface of airways is covered with a liquid film, called airway surface liquid (ASL). The diameter of the airways decreases gradually from the trachea to the alveolar sacs, and this can be  characterised to leading order by the correlation $a_j=a_02^{(-j/3)}$ for the first fourteen  generations, where $a_0$ is the diameter of the trachea and $a_j$ is the diameter of the airway at $\textnormal{j}^{\textnormal{th}}$ generation \citep{Weibel1962}. The lung can be divided into two parts, which are the conducting and the respiratory zones. The first 16 generations are the conducting zone that includes the trachea, the bronchi, the bronchioles and the terminal bronchioles \citep{Grotberg2011}. This section of the lungs is responsible for the transport of air to the respiratory zone, so it forms the anatomic dead space \citep{west2012respiratory}. After the terminal bronchioles, the respiratory zone starts with respiratory bronchioles and ends with the alveolar sacs $(j=23)$ \citep{Grotberg2011}. The respiratory zone is the part of the lung, where gas exchange occurs in the alveoli.

Airway closure can occur due to a Plateau-Rayleigh instability when the ASL is too thick. This phenomenon can be observed in healthy conditions whenever the lung volume is low, and in diseased conditions, such as asthma, pulmonary edema, and respiratory distress syndrome \citep{halpern2008liquid}. \citet{Gauglitz1988} stated that for a clean rigid pipe, lined with a single-layer liquid, the critical thickness to observe coalescence is $h_c^*/a^*\approx0.12$, where $h_c^*$ is the critical thickness of the liquid film and $a^*$ is the radius of the pipe. \citet{Halpern1992,Halpern1993} investigated the stability of a liquid film lining an airway tube by considering the wall elasticity and surfactants. They concluded that the increasing wall compliance decreases $h_c^*$, while surfactants increase it. Later, \citet{Halpern2003} analysed the effects of an oscillatory core flow on more viscous liquid film coating a rigid cylinder. They found that the core flow can avoid liquid plug formation in a non-linear fashion by spreading back and forth the ASL. That is what they termed "reversing butter knife effect". Additionally, a \textit{compliant collapse}, can also be observed in certain situations, where fluid-elastic instabilities arise \citep{Heil2008}. If there is a liquid bridge formation without a structural collapse, it is called \textit{film collapse} \citep{Kamm1989}.

For about the first 16 generations, ASL is a bilayer, where the sublayer is serous (serum) and the top layer is mucus \citep{grotberg2001respiratory}. Although the serum predominantly shows Newtonian characteristics, airway mucus is often reported as a highly non-Newtonian substance \citep{Girod1992, Lai2009, Cone2009}. About 90-95\% of airway mucus comprises of water, and it is followed by high-molecular mucin glycoproteins with 2-5\% \citep{spagnolie2015complex}. It, additionally, includes fractional amounts of lipids, salts, DNA and cell debris \citep{spagnolie2015complex}. This mixture of materials gives airway mucus its structured form, and thus its non-Newtonian features \citep{Lai2009}.

The solid content of airway mucus is responsible for its viscoelastic/viscoplastic characteristics. It was shown by \citet{Hill2014} that these characteristics are directly proportional to the amount of solid in mucus and consequently pulmonary diseases. For example, while solid concentration for a normal pulmonary mucus is around 2 wt \%, for a sample from a chronic obstructive pulmonary disease (COPD) patient, it is almost 4\%, and it can even go as high as 8\% for a cystic fibrosis (CF) patient \citep{Hill2014}. Also, \citet{lafforgue2018rheological} correlated the rise in the solid concentration to the increase of viscoplastic and shear-thinning features by fitting a Herschel-Bulkley model to their steady-state data. This abnormal increase in the viscoelastic/viscoplastic characteristics is usually followed by an increase in viscosity levels as well, and mucociliary clearance can eventually be compromised \citep{Williams2006}. A direct result of this is the growth in the number of bacterial pathogens \citep{Lai2009}. Additionally, COPD, CF and asthma can cause a mucus hyper-secretion and obstruct the airway in lethal degrees \citep{Williams2006}.

Rheology of the airway mucus samples obtained from healthy \citep{patarin2020rheological, schuster2013nanoparticle} and diseased subjects \citep{Dawson2003, nettle2018linear, nielsen2004elastic, patarin2020rheological} has also been studied extensively in the literature. These investigations provided strong indications about viscoelasticity, viscoplasticity and shear-thinning features, and thus elastoviscoplasticity of the pulmonary mucus. Elastoviscoplastic (EVP) fluids can be seen in many areas of our lives from industry to nature. This fluid behaviour involves a critical stress (yield stress), above which material starts to flow, and below which the behaviour of the material is similar to that of an elastic solid \citep{Saramito2007}. The existence of the yield stress causes a singularity in the deformation of the material, and methods such as viscosity regularisation, the augmented Lagrangian method and mapping of the yield surface to a fixed boundary have been used to overcome the resulting difficulties in modelling. \citep{Fraggedakis2016evpcomparison}.

Although simplified viscoplastic constitutive models have been extensively used in the literature, behaviours of most of the real fluids cannot be described solely using variants of these models. For example, the experiments on well-charaterised yield-stress fluids \citep{gueslin2006flow, putz2008settling, Holenberg2013} showed that the fore-aft symmetry, which had been estimated using classical viscoplastic model assumptions \citep{beris1985creeping}, was lost, giving rise to the formation of a "negative wake" \citep{Fraggedakis2016sphericalparticle}. This demonstrated that yield-stress fluids are rather more complex because of their thixotropy and elasticity \citep{Holenberg2013}.

In order to include viscoelastic effects that is observed in some viscoplastic fluids, \citet{desouzamendes2007dimensionless} suggested a modification of lower-convected Jeffreys liquid, where viscosity, relaxation time and retardation time are functions of the deformation rate. When the material yields, the proposed constitutive equation reduces to the generalised Newtonian liquid constitutive equation. Otherwise, the model reduces to the Jeffreys liquid constitutive equation. \citet{Benito2008} developed a fully tensorial continuous framework to describe the behaviour of soft materials, which deform substantially before yielding. After yielding, the material flows as a viscoelastic fluid. In their work, where they studied the oscillatory pipe flow of a Carbopol solution, \citet{Park2010} both carried out experiments to obtain the velocity fields and compared the results with the computational solutions based on the EVP model that they proposed. The model is made of elastic springs connected to a regularised Bingham model, and agreed well with the experimental findings. \citet{Belblidia2011} proposed another constitutive law to take the elastoviscoplasticity into account. In their model, they built upon the model of \citet{Papanastasiou1987}, and to include the viscoleastic effects, they used the Oldroyd-B model \citep{oldroyd1950formulation}. 

Based on the thermodynamic theory, \citet{Saramito2007} proposed a 3-D constitutive model, which combines the Bingham viscoplastic and the Oldroyd-B viscoelastic models. Accordingly, the behaviour of the material is a viscoelastic solid before yielding, and when the stress exceeds a critical level, the material behaves as a viscoelastic fluid. Here, the von Mises criterion is used to monitor the yielding. Later, this model is improved to include the shear-thinning behaviour by combining the Oldroyd viscoelastic and Herschel-Bulkley models \citep{Saramito2009}. Recently, \citet{Fraggedakis2016evpcomparison} compared five constitutive models (three variations of \citet{Saramito2007}, \citet{Park2010} and \citet{Belblidia2011}) by performing series of tests, such as simple-shear, uniaxial elongation, and large amplitude oscillatory tests and found that the Saramito variants outperformed the other two models.

\begin{comment}
Large amplitude oscillatory shear (LAOS) tests have been designed to explore the non-linear behaviours of materials for many years \citep{hyun2011review}. These tests can be carried out in two ways, stress-controlled (LAOStress) and strain-controlled (LAOStrain) \citep{dimitriou2013describing}. Both methods basically depend on the stress or strain decomposition based on Fourier series. In the non-linar region of the test, $G^{'}$, elastic or storage modulus, and $G^{''}$, viscous or loss modulus, lose their exact meaning because the response is no longer sinusoidal due to higher harmonics \citep{giacomin1993large}. Therefore, a Fourier analysis is normally carried out on the obtained response to analyse the results \citep{wilhelm2002fourier}. This analysis reveals that as the strain amplitude increases, odd harmonics of the response start to be seen, and they, in fact, increase their magnitude in large strain or stress amplitudes. As an alternative to Fourier transform (FT) definitions, there are also other definitions of the viscoelastic moduli in the nonlinear region, such as the ones obtained by \citet{ewoldt2008new} using Chebyshev stress decomposition. Today, most of the commercial rheometers report the first-harmonic moduli ($G^{'}=G_1^{'}$ and $G^{''}=G_1^{''}$) \citep{hyun2011review}. The same convention is also used in this paper, unless otherwise is stated.
\end{comment}

Liquid plug formation, propagation and rupture exert potentially lethal mechanical stresses to the airway wall, where airway epithelium lies \citep{Bilek2003, Kay2004, Huh2007, Tavana2011}. The pre-coalescence dynamics of the airway closure in a rigid tube lined by a single-layer Newtonian fluid has been studied experimentally and numerically by \citet{Bian2010} and \citet{Tai2011}, respectively. They both concluded that mechanical stresses may reach to the levels marked dangerous for the airway epithelium by \citet{Bilek2003} and \citet{Huh2007} during the plug formation. Later, \citet{Romano2019} investigated the whole closure process, including the post-closure dynamics, by modelling the ASL as a single-layer Newtonian liquid film. They deduced that stress peaks occurring just after the coalescence, during \textit{bi-frontal plug growth}, are responsible for high mechanical stresses exerted on the pulmonary epithelium. The effects of the complex characteristics of the airway mucus on pre-coalescence dynamics were taken into account by \citet{Halpern2010} using lubrication approximations. In their study, they modelled the ASL as a one-layer Oldroyd-B fluid, and analysed the effect of the Weissenberg number, $Wi$, on the growth rate of the instabilities and the wall shear stress levels. \citet{Romano2021} recently considered viscoelastic effects in a similar problem by using Oldroyd-B and FENE-CR models. They showed that mucus viscoelasticity is responsible for the second peak of the wall shear stress occurring after the coalescence, and this secondary peak can be as extreme as the first one for high Laplace and Weissenberg numbers in a physiological range. Instability of an axisymmetric layer of viscoplastic Bingham liquid coating the interior of a rigid tube is studied by \citet{shemilt2022surface}. This model represents the airway and takes into account the yield stress of mucus. Using long-wave theory, they derived an evolution equation for the thickness of the liquid layer. They found that as the capillary Bingham number increases, the critical layer thickness required to form a liquid plug also increases. Recently, \citet{erken2022capillary} studied this problem in a two-layer setting, where both layers were Newtonian. The main findings were the enhanced instability of the system leading to a sooner closure and the damping of the stresses, both of which were related to the existing of the bottom (serous) layer. Moreover, the non-Newtonian effects of mucus have been studied in plug propagation and rupture both numerically and experimentally \citep{zamankhan2012steady, hu2015microfluidic, hu2020effects, zamankhan2018steady, bahrani2022propagation}, and it has been reported that these features should be considered in airway models.

Airway mucus is a very complex material and exhibits a wide range of non-Newtonian characteristics, such as viscoelasticity, viscoplasticity, shear-thinning, and thixotropy \citep{Girod1992}. Therefore, a mucus model that incorporates these features is needed to obtain more realistic results for the airway closure. In this paper, the effects of the non-Newtonian characteristics of pulmonary mucus on the airway closure problem have been investigated using the Saramito-HB model \citep{Saramito2009}. This constitutive law is a combination of Oldroyd-B and Herschel-Bulkley models, with a power-law index $n>0$. The model parameters are determined by following a parameter fitting procedure similar to that of \citet{Fraggedakis2016sphericalparticle} using the experimental results of \citet{patarin2020rheological} for the healthy, asthma, COPD, and CF airway mucus. To compare different pathological conditions of the pulmonary mucus in different settings, extensive simulations are performed for varying surface tension and initial liquid layer thickness. Here, we study the airway closure problem in a single-layer setting in order to isolate the effects of the EVP features of the mucus on the closure time and the wall mechanical stresses. The system is further simplified by neglecting the wall deformation and the surfactants. A similar framework was studied by \citet{Romano2021}, where the liquid lining was a viscoelastic fluid but they considered only the healty conditions. Here, we also include viscoplastic and shear-thinning properties of mucus by utilising Saramito-HB model. To the authors' knowledge, effects of the mucus elastoviscoplasticity on airway closure have never been studied so far.

The rest of the paper is organised as follows. Mathematical formulation and numerical method are explained in \S\ref{sec:formulation}. Then, the problem specifications, such as boundary conditions and parameter intervals, are described in \S\ref{sec:problemStatement}. \S\ref{sec:resultsAndDiscussion} presents the rheological fitting and simulation results by specifically focusing on the effects of different pathological conditions. Finally, the summary and the conclusions of the study are given in \S\ref{sec:summaryAndConclusions}.

%%%%%%%%%%%%%%%%%%%%%%%%%%%%%%%%%%%%%%%%%%%%%
%%%%%%%%%%%%%%%%%%%%%%%%%%%%%%%%%%%%%%%%%%%%%
\section{Formulation and numerical method}\label{sec:formulation}
The governing equations are described in the context of the finite-difference/front-tracking method \citep{Unverdi1992}. Using a one-field formulation, a single set of incompressible momentum and continuity equations is written in the whole computational domain. The interfacial effects are represented as a body force in the momentum equation, and the jumps in the material properties for the different phases are accounted for using an indicator (color) function. The equations are solved in their dimensional forms denoted by superscript `$^{*}$'. In the front-tracking framework, the momentum and continuity equations yield

\begin{equation} %2.1
\begin{aligned}
\frac{ \partial{\rho^{*} \boldsymbol{u}^*} }{ \partial{t^*} } + \boldsymbol{\nabla}^*\boldsymbol{\cdot} \left(\rho^*\boldsymbol{u}^*\boldsymbol{u}^*\right) = &-\boldsymbol{\nabla}^*p^* + \boldsymbol{\nabla}^*\boldsymbol{\cdot}\mu_{l,S}^*\left(\boldsymbol{\nabla}^*\boldsymbol{u}^* + \boldsymbol{\nabla}^*{\boldsymbol{u}^*}^T\right) + \boldsymbol{\nabla}^*{\cdot}\boldsymbol{\tau}^* \\
&+ \int_{A^*}{\sigma^*\kappa^*\boldsymbol{n}\delta\left(\boldsymbol{x}^* - \boldsymbol{x}_f^*\right)dA^*},
\end{aligned}
\label{eqn:momentum}
\end{equation}

\begin{equation} %2.2
\boldsymbol{\nabla}^*\boldsymbol{\cdot}\boldsymbol{u}^* = 0,
\label{eqn:continuity}
\end{equation}
where $t^*$ is the time, $\boldsymbol{u}^*$ is the velocity vector, $p^*$ is the pressure field, $\rho^*$ and $\mu_{l,S}^*$ are the discontinuous density and solvent viscosity fields, respectively, and $\boldsymbol{\tau}^*$ represents the extra stress tensor. Note that $\mu_{l}^*=\mu_{l,S}^* + \mu_{l,P}^*$, where $\mu_{l,P}^*$ and $\mu_{l}^*$ are the polymer and total viscosities of the liquid layer, respectively. The effect of the surface tension is represented as a body force in the last term on the right-hand side of equation~(\ref{eqn:momentum}), where $\sigma^*$ is the surface tension coefficient, $\kappa^*$ is twice the mean curvature and $\boldsymbol{n}$ is a unit vector normal to the interface, and $A^*$ is the surface area. The surface tension acts only on the interface as indicated by the Dirac delta function $\delta$, whose arguments $\boldsymbol{x^*}$ and $\boldsymbol{x}_f^*$ are the points at which the equation is evaluated and the point at the interface, respectively. The gravitational effects are negligible within the asymptotic limit of a small Bond number, i.e., $\text{Bo}= g^* {a^*}^2\Delta\rho^*/\sigma^* \ll 1$, where $g^*$ is the gravitational acceleration, $\Delta\rho^*$ is the difference between mucus and air densities. Since the airway flow of interest in the present study falls in such a regime, the gravitational effects are neglected in equation (\ref{eqn:momentum}).

The liquid layer is represented as a Saramito-HB fluid \citep{Saramito2009}. The EVP equations are solved using the log-conformation method \citep{Izbassarov2015}. The extra stresses appearing in equation (\ref{eqn:momentum}) are related to the conformation tensor, $\boldsymbol{B}$, that evolves by
\begin{equation} %2.3
\left( \frac{\partial\boldsymbol{B}}{\partial t^*} + \boldsymbol{u^*}\cdot\boldsymbol{\nabla B} - \boldsymbol{B}\cdot\boldsymbol{\nabla u^*} - \boldsymbol{\nabla u^*}^T\cdot\boldsymbol{B} \right) = \frac{F}{\Lambda^*}\left( \boldsymbol{I} - \boldsymbol{B} \right),
\label{eqn:transport}
\end{equation}
where $\Lambda^*$ is the relaxation time. In the Saramito-HB model, the relaxation time and the polymeric viscosity are given as $\Lambda^*=\mu_{l,P}^*/G^*$ and $\mu_{l,P}^*=K^*\left( L^*/U^* \right)^{(1-n)}$, respectively, where $L^*$, $U^*$, $G^*$, $n$ and $K^*$ are characteristic length and velocity scales, elastic modulus, power-law index and consistency parameter, respectively \citep{Saramito2009}. The Saramito-HB model can be represented by setting the parameters of equation (\ref{eqn:transport}) to $\boldsymbol{B}=\boldsymbol{\tau}^*\lambda^*/\mu_{l,P}^*+\boldsymbol{I}$ and $F/\Lambda^*=G^*\textnormal{max}\left[ 0,\frac{\mid{\boldsymbol{\tau}^*}^d\mid - \tau_y^*}{K^*\mid{\boldsymbol{\tau}^*}^d\mid^n} \right]^{\frac{1}{n}}$ \citep{Izbassarov2015, Izbassarov2018}, where $\tau_y^*$ is the yield stress and $\boldsymbol{\tau^*}^d$ is the deviatoric part of the stress tensor, and its magnitude is given as

\begin{equation} %2.4
\mid {\boldsymbol{\tau}^*}^d \mid = \left( \frac{1}{2}{\tau^*}^d_{ij}{\tau^*}^d_{ij} \right)^\frac{1}{2}.
\label{eqn:deviatoricstress}
\end{equation}

Once the conformation tensor is obtained from equation \ref{eqn:transport}, the extra stress tensor is then computed as $\boldsymbol{\tau}^* = \frac{\mu_{l,P}^*}{\Lambda^*}\left( \boldsymbol{B} - \boldsymbol{I} \right)$. Also, it is assumed that the material properties remain constant following a fluid particle, i.e.,
\begin{equation} %2.5
\begin{aligned}
\frac{ D\rho^*}{ D t^* }=0&,&\hspace{0.7cm} \frac{ D\mu_{l,S}^*}{ D t^* }=0&,&\hspace{0.7cm} \frac{ D G^*}{ D t^* }=0&,&\hspace{0.7cm}\\ \frac{ D \tau_y^*}{ D t^* }=0&,&\hspace{0.7cm} \frac{ D n}{ D t^* }=0&,&\hspace{0.7cm} \frac{ D K^*}{ D t^* }=0&,&
\end{aligned}
\label{eqn:materialprop-derivative}
\end{equation}
where $D/Dt^* = (\partial/\partial t^*) + \boldsymbol{u}^* \boldsymbol{\cdot} \boldsymbol{\nabla}^*$ is the material derivative. The material properties vary discontinuously across the interfaces and are given, for example, for density $\rho$ by
\begin{equation} %2.6
  \rho^* = \rho_g^* I(r,z,t) + \rho_l^*\left[1-I(r,z,t)\right]
     \label{eqn:materialprop-distribution}
\end{equation}
where the subscripts "$g$" and "$l$" denote the properties of the air core and the liquid layer, respectively, and $I$ is the indicator function having the values $I=0$ in the liquid layer and $I=1$ in the air core. All the other material properties, given in equation (\ref{eqn:materialprop-derivative}), are distributed in the same way across the computational domain depending on the value of $I$.

The flow equations are written and solved in the context of the front-tracking/finite-difference method \citep{Unverdi1992}. This method contains two grids that are a stationary staggered Eulerian grid, where the flow equations are solved to obtain the velocity, pressure and extra stress fields, and a Lagrangian grid, which is formed by marker points. The piece of the interface between two neighbouring marker points forms a front element. Both the marker points and the front elements are connected to form the air-liquid interface. The material properties are distributed according to the location of the interface (front) at the beginning of each time step according to equation (\ref{eqn:materialprop-distribution}). To compute the surface tension at the centroids of the front elements, a third-order Legendre polynomial fit is used. Then, the surface tension is distributed smoothly onto Eulerian grid points to be added to the momentum equations as a body force to account for the interfacial effects. At each time step, the local flow velocity of the front is interpolated from the Eulerian grid, and the front is moved accordingly. The communication between the Eulerian and the Lagrangian grids are accomplished by using Peskin's cosine distribution function \citep{Peskin1977}.

To approximate the spatial derivatives, central differences are used, except for the convective term in equation \ref{eqn:transport}, where a $5^{\rm th}$ order WENO-Z method is used. Time integration is accomplished by the projection method developed by \citet{Chorin1968}. The method is first-order accurate in time, but a second-order accuracy can easily be achieved by a predictor-corrector scheme as described by Tryggvason et al.\cite{Tryggvason2001}. However, Muradoglu et al.\cite{Muradoglu2019} note that for the first-order method in use, the time-stepping error is smaller compared to the spatial error due to the tight restrictions of the stability condition on the time step for the flow of interest in this study. Hence, the first-order method is here employed.

The extra stress is calculated at each time step and added to the momentum equations to represent the EVP effects \citep{Izbassarov2018}. To that end, the generic transport equation (\ref{eqn:transport}) is solved by changing its parameters for the Saramito-HB model in use. For a detailed explanation of the numerical procedure on how this equation is solved and extra stresses are handled, the reader is referred to \citet{Izbassarov2015}. 

The indicator function is calculated according to the location of the front at the beginning of each time step as explained by \citep{Tryggvason2001}. For this purpose, a separable Poisson equation, resulting from the divergence of the unit magnitude jumps calculated at the centres of the front elements and distributed onto the neighbouring Eulerian grid cells, is solved. Once the indicator function is computed in the whole domain, the material properties are updated according to equation (\ref{eqn:materialprop-distribution}). Afterwards, the integration is carried out according to these updated properties to obtain the velocity and pressure fields. 

The density of marker points in the Lagrangian grid, thus the size of the front elements, is monitored at each time step to prevent numerical inaccuracies and instabilities. Too coarse grid can cause poor resolution of the interface, whereas too dense grid can cause unwanted wiggles. Therefore, the front elements are kept between pre-specified minimum and maximum sizes by splitting large elements by adding new marker points or deleting small elements. During this restructuring of the front, the curvature is considered by using a third order Legendre interpolation to preserve the smoothness of the interface.

In the front-tracking method, the marker points are explicitly tracked, so the topological change must be implemented by changing the connectivity of the marker points in an appropriate way \citet{Tryggvason2001}. The procedure suggested by Olgac et al.\cite{Olgac2006} is used in the present study to handle the topological change. Thus, the minimum distance between the interface and the symmetry axis is monitored. When this falls below a pre-specified minimum limit, $l_{\rm th}$, the front element that is closest to the symmetry axis is removed, and the interface is connected to the symmetry axis. The effects of this threshold value, $l_{\rm th}$, on the results are checked, and it is found that as long as $l_{\rm th}$ is of the order of the Eulerian grid size, the results are uninfluenced.

A detailed information about the front-tracking method can be found in \citet{Unverdi1992}, \citet{Tryggvason2001}, and \citet{tryggvason2011direct}. The method has already been validated successfully for an airway closure problem with a Newtonian single-layer case against the results of \citet{Romano2019}, obtained by using volume-of-fluid (VOF) method and implemented in \textit{basilik} package \citep[][http://basilisk.fr]{Popinet2014}. The implementation of the Saramito-HB model is first validated against the analytical solution of a single-phase laminar pipe flow given by \citet{Chaparian2019}. Saramito-HB model reduces to Oldroyd-B model when $n=1$ and $\tau_y^*=0$ \citep{Saramito2009}, so the implementation of the non-Newtonian model is further validated against the results of \citet{Romano2021} for the airway closure with a single-layer Oldroyd-B liquid. Although not included here, the results were found to be in good agreement with those of \citet{Romano2021}

%%%%%%%%%%%%%%%%%%%%%%%%%%%%%%%%%%%%%%%%%%%%%
%%%%%%%%%%%%%%%%%%%%%%%%%%%%%%%%%%%%%%%%%%%%%
\section{Problem statement}\label{sec:problemStatement}
\begin{figure}
  \centerline{\includegraphics[trim={0cm 0cm 0cm 0cm},clip,scale=0.63]{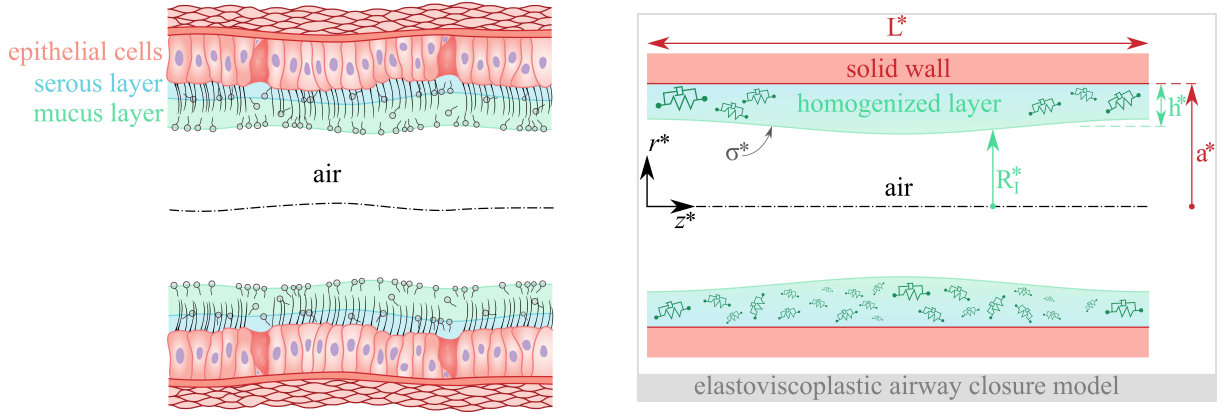}}
  \caption{ (a) The schematic illustration of a part of an airway. (b) The illustration of the computational domain. The axisymmetric rigid airway wall is lined with the EVP liquid layer, and the liquid layer is surrounded by the air core inside. The length and the radius of the airway lumen is $L_z^*$ and $a^*$, respectively. The surface tension of the air-liquid interface is $\sigma^*$. The radial location of the interface, which is perturbed from its initial state to initiate the instability, is shown by $R_I^*$. }
\label{fig:geometry}
\end{figure}

A schematic of the problem is given in figure \ref{fig:geometry}. The airway lumen, whose length and radius are $L_z^*$ and $a^*$, respectively, is axisymmetric around the centreline, and its wall is rigid. Here, the air core is modelled as a Newtonian fluid with constant properties, which are denoted by the subscript "$g$". On the other hand, the liquid layer is modelled as an EVP fluid, and its properties are denoted by subscript "$l$". The surface tension at the air-liquid interface, $\sigma^*$, is assumed to be constant as the other material properties of the air and the liquid film. The computational domain is periodic at $z^*=0$ and $z^*=L_z^*$, and no-slip boundary conditions are applied at the rigid wall. The liquid layer is perturbed from its initial location, $h^*$, to initiate the instability, and the radial location of the interface is given by

\begin{equation}
    \begin{aligned}
        r^* & = R_I^* = a^* - h^*\left[1-0.1 \hspace{0.1cm} \times \hspace{0.1cm} cos\left(2 \pi z^* / L_z^*\right)\right], \\
    \end{aligned}
    \label{eqn:InitialPerturbation-ol}
\end{equation}where $R_I^*$ is the radial location of the air-liquid interface, and $z^*$ and $r^*$ are the axial and radial coordinates, respectively.

The equations are solved in their dimensional form, as mentioned above, and the results are presented in terms of non-dimensional groups, by making use of a capillary scaling, i.e., length, time, velocity and stresses are non-dimensionalised by $a^*$, $\mu_{l,S}^*a^*/\sigma^*$, $\sigma^*/\mu_{l,S}^*$ and $\sigma^*/a^*$, respectively. The resulting non-dimensional parameters can be summarised as

\begin{equation}
\begin{aligned}
 La&=\frac{\rho_l^* \sigma^* a^*}{{\mu_{l,S}^*}^2}, \hspace{0.7cm} &Bi=&\tau_y=\frac{\tau_y^*a^*}{\sigma^*}, \hspace{0.7cm} &G=&\frac{G^*a^*}{\sigma^*}, \hspace{0.7cm} &Wi=&\frac{\Lambda^* \sigma^*}{a^* \mu_{l,S}^*},&\\ 
 \rho&=\frac{\rho_g^*}{\rho_l^*}, &\mu=&\frac{\mu_g^*}{\mu_{l,S}^*}, &\epsilon=&\frac{h^*}{a^*}, 
&\lambda=&\frac{L_z^*}{a^*},
\end{aligned}
\label{eqn:NondimensionalParameters-ol}
\end{equation}
where $La$, $\tau_y$, $G$, $\rho$, $\mu$, $\epsilon$, and $\lambda$ denote Laplace number, non-dimensional yield stress, non-dimensional elastic modulus, gas-to-liquid density ratio, gas-to-liquid viscosity ratio, non-dimensional initial film thickness, and airway tube length-to-radius ratio, respectively. $Bi$ is the Bingham number and $Wi$ is the Weissenberg number. Also, $\mu_{l,S}^*$ and $\mu_g^*$ are the liquid solvent viscosity and the total gas viscosity, respectively, and $\mu_l^*=\mu_{l,S}^*+\mu_{l,P}^*$, where $\mu_{l}^*$ is the total liquid viscosity and $\mu_{l,P}^*$ is the polymeric viscosity. In the Saramito-HB model, the relaxation time is defined as $\Lambda^*=\mu_{l,P}^*/G^*$. Here $\mu_{l,P}^*=K^*\left(\frac{L^*}{U^*}\right)^{1-n}$, where $L^*$ is the length scale and $U^*$ is the velocity scale.

The parameter ranges are determined to represent the ninth-to-tenth generation of a typical adult human lung. The airways can be compared to branching tubular trees, and the radii of these tubes decrease at each generation \citep{Weibel1962}. The airway closure starts to be seen after ninth or tenth generation because airway radii are not small enough in the earlier ones \citep{Breatnach1984, Burger1968}. Thus, here, the airway radius is taken as $a^*=0.065$ \si{cm} \citep{crystal1997lung}. Also, it should be noted that a typical airway lumen has a length-to-radius ratio of $\lambda=6$ \citep{Kitaoka1999}, so this value is used throughout this paper. 

For a Newtonian one-layer liquid lining a clean rigid pipe, the liquid plug formation starts to occur when $h_c^*/a^*\ge0.12$, where $h_c^*/a^*$ is the critical initial film thickness \citep{Gauglitz1988}. Accordingly, in this study, the initial non-dimensional film thickness, $\epsilon$, is varied in the range of $0.25\le\epsilon\le0.35$ to describe different intensities of mucus hypersecretion.

\citet{Romano2019} stated that the density of the single-layer liquid can be taken as $\rho_l^*=1000$ \si{kg/m^3}. Also, the solvent viscosity of the liquid layer is fixed at $\mu_{l,S}^*=0.013$ \si{Pa \cdot s} \citep{Tai2011}. On the other hand, to represent the surfactant-deficient conditions, three different surface tension values are used, $\sigma^*=0.026$ \si{N/m}, $\sigma^*=0.052$ \si{N/m} and $\sigma^*=0.078$ \si{N/m} corresponding to $La=100$, $La=200$ and $La=300$, respectively \citep{Moriarty1999, Schurch1990}. 

The properties of the airway mucus are determined for the healthy, asthma, CF and COPD cases using the experimental data of \citet{patarin2020rheological} to examine the effects of the pathological conditions on the airway closure phenomenon. The EVP parameters of the liquid layer for these conditions are extracted using a non-linear regression, as explained in detail in section \ref{subsec:rheologicalFitting}.

After the healthy, asthma, COPD and CF conditions are compared, a parametric study is carried out to see the effects of individual parameters of the Saramito-HB model. For this purpose, the healthy case is taken as the baseline, and the elastic modulus $G^*$, the yield stress $\tau_y^*$ and the shear-thinning index $n$ are varied in a physiologically meaningful range. \citet{hu2015microfluidic} stated that $G^*$ can reach up to $G^*\approx200$ \si{Pa} depending on the pathological conditions and the angular frequency. In another study, they studied the plug rupture problem numerically by modelling the liquid film as a Herschel-Bulkley fluid \citep{hu2020effects}, where they varied $\tau_y^*$ up to the extreme conditions $\tau_y^*=100$ \si{Pa}. Based on the work of \citet{lafforgue2018rheological}, $n$ can also vary depending on the solid concentration of mucus. By fitting a Herchel-Bulkley model to their experimental data on a mucus simulant proposed by \citet{zahm1991role}, they found that $0.37 \le n \le 0.78$. Therefore, in our parametric study, which is omitted in this paper for brevity, the elastic modulus $G^*$, yield stress $\tau_y^*$ and power-law index $n$ are studied in the ranges of $G^* \in [0.094,100]$ \si{Pa}, $\tau_y^* \in [0.04,100]$ \si{Pa} and $n \in [0.4,1.0]$. In terms of the non-dimensional quantities, these lead to $G \in [0.0013,2.5]$, $\tau_y \in [0.01,2.5]$ and $n \in [0.4,1.0]$. The rheological fitting for these parameters (explained in section \ref{subsec:rheologicalFitting}) is performed within these ranges as well.

%%%%%%%%%%%%%%%%%%%%%%%%%%%%%%%%%%%%%%%%%%%%%
%%%%%%%%%%%%%%%%%%%%%%%%%%%%%%%%%%%%%%%%%%%%%
\section{Results and discussion}\label{sec:resultsAndDiscussion}
Firstly, the effects of pathological conditions of mucus are studied. In \S\ref{subsec:rheologicalFitting}, a parameter fitting algorithm (similar to Fraggedakis et al. \cite{Fraggedakis2016sphericalparticle}) is followed to obtain the parameters of the Saramito-HB model for healthy, asthma, COPD and CF mucus based on the experimental data of \citet{patarin2020rheological}. Afterwards, these mucus states are compared by varying $\epsilon$ and $La$ to represent different intensities of mucus hypersecretion and surfactant deficiency. Finally, a parametric study is performed on the healthy mucus to see the individual effects of the parameters.

The computational domain is given in figure \ref{fig:geometry}, and it has an axial length of $L_z^*=6a^*$ and a radial length of $a^*$. A uniform tensor-product structured grid is used in all simulations carried out. Grid convergence is checked, and it is found that a stretched Cartesian grid of $96\times576$ is enough to reduce the spatial errors below 4\%, for the wall shear stress excursion, $\Delta \tau_w =$ max($\tau_w$) - min($\tau_w$), the wall pressure excursion, $\Delta p_w =$ max($p_w$) - min($p_w$), and the minimum core radius, $R_{min}$. The grid is stretched so that the radial grid size, $\Delta{r}$, is three times smaller at the wall compared to the a grid near the centerline.

%%%%%%%%%%%%%%%%%%%%%%%%%%%%%%%%%%%%%%%%%%%%%%%%%%%%%%%%%%
\subsection{Determination of rheological properties}\label{subsec:rheologicalFitting}
In this section, the Saramito-HB model is fitted to the experimental data provided by \citet{patarin2020rheological}, and the model parameters for the Saramito-HB model are obtained. In their set of experiments, they collected mucus samples from healthy, asthma, COPD and CF subjects, and analysed them in a quite large strain amplitude interval, by fixing the frequency to 0.6 \si{Hz}. Moreover, they presented the evolution of elastic modulus ($G^{'}$) and viscous modulus ($G^{''}$) by varying strain amplitude in a large interval, and concluded that these four different conditions have distinct rheological behaviours.

Firstly, the Saramito-HB constitutive law is written in its 1-D form as Fraggedakis et al.\cite{Fraggedakis2016evpcomparison} suggested

\begin{equation}
\begin{aligned}
 \frac{1}{G^*}\overset{\triangledown}{\boldsymbol{\tau}^*} + max\left[ 0,\frac{\mid{\boldsymbol{\tau}^*}^d\mid - \tau_y^*}{K^*\mid{\boldsymbol{\tau}^*}^d\mid^n} \right]^{\frac{1}{n}} \boldsymbol{\tau}^* = 2 \boldsymbol{D}^*
\end{aligned}
\label{eqn:saramitohb-1D}
\end{equation}  
where $G^*$, $\tau_y^*$, $n$ and $K^*$ are the elastic modulus, the yield stress, the power-law index and the consistency parameter, respectively. $\mid{\boldsymbol{\tau}^*}^d\mid$ is the magnitude of the deviatoric part of the extra stress tensor, $\boldsymbol{\tau}^*$, and its definition is given in (\ref{eqn:deviatoricstress}). The symbol "$\triangledown$" above $\boldsymbol{\tau}^*$ denotes the upper-convected Maxwell derivative, and finally, $\boldsymbol{D}^*$ is the deformation tensor, which is defined as $\boldsymbol{D}^*=\frac{1}{2}\left[ \left(\nabla^*\boldsymbol{u}^*\right) + \left(\nabla^*\boldsymbol{u}^*\right)^T \right]$. According to equation (\ref{eqn:saramitohb-1D}), the material exhibits a shear-thinning behaviour when $0<n<1$ and an unusual shear-thickening behavior when $n>1$ \citep{Saramito2009}.

Then, the equation is solved according to its simple shear solution. In LAOStrain, strain-controlled Large amplitude oscillatory shear (LAOS) test \citep{dimitriou2013describing, hyun2011review}, the input strain is given by 

\begin{equation}
\gamma=\gamma_o sin(\omega^*t^*)
\label{eqn:sinusoidalstrain}
\end{equation}
where $\omega^*$ is the input angular frequency. However, it should be noted that the Saramito-HB model does not take the strain as an input, so the strain rate is defined as $\dot{\gamma}=\partial_t\gamma$ and it is computed by taking the time derivative of the strain input. The input velocity field is defined as the simple shear flow

\begin{equation}
\boldsymbol{u}^*=\left( \dot{\gamma}^*(t^*)y^*,0,0 \right)
\label{eqn:velocityfield}
\end{equation}

\begin{figure}
  \centerline{\includegraphics[trim={0.0cm 0.0cm 0.0cm 0.0cm},clip,scale=0.90]{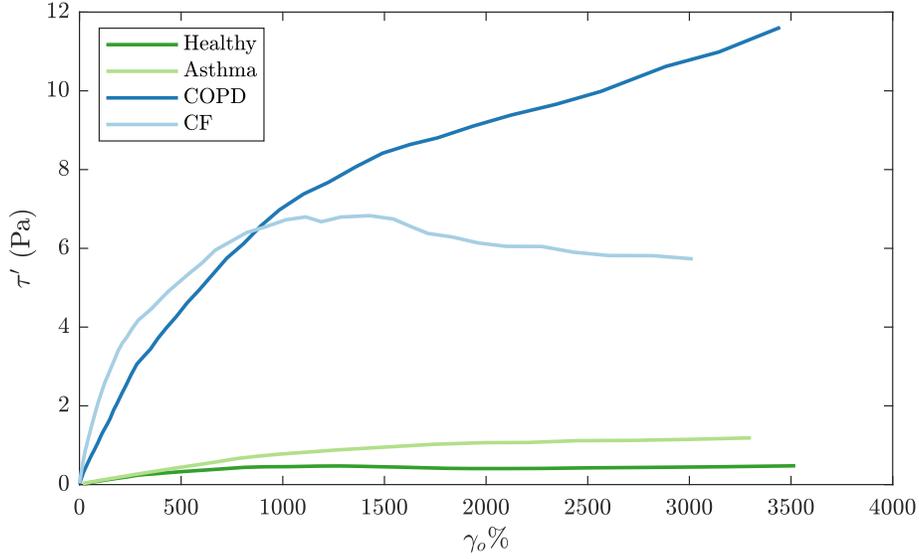}}
  \caption{ The in-phase stress component vs. strain amplitude, $\gamma_o$ for healthy, asthma, COPD and CF mucus. }
\label{fig:inphase-stress}
\end{figure}

Before starting with the non-linear regression, the yield stress, $\tau_y^*$, of the material is obtained by a method proposed by Yang et al.\cite{yang1986some}. This method suggests that the in-phase stress component is given by $\tau^{'}=G^{'}\gamma_o$. When this stress component is plotted against the varying strain amplitudes, the maximum value that the in-phase stress component attains is the yield stress of the material, i.e., the stress, at which structural breakdown occurs. The variation of $\tau^{'}$ by $\gamma_o\%$ is given in figure \ref{fig:inphase-stress}. The yield stresses found following this method are given in table \ref{tab:rheologicalFitting}. After the yield stresses are obtained for the four types of mucus samples, then the fitting is done for the remaining 3 parameters ($G^*$, $n$ and $K^*$). Note that for the COPD case, there is not a conclusive yield stress from this method, so for this condition of mucus, the yield stress value is taken from table 2 of \citet{patarin2020rheological} ($\sigma_c$ for COPD, spontaneous case, where no induction is necessary for patients to expectorate sputum).

For the fitting procedure, a cost function is used to determine the best possible fitting to the experimental data. As the cost function, a modified version of Fraggedakis et al.\cite{Fraggedakis2016sphericalparticle} is used, which is

\begin{equation}
Cost(G^*,n,K^*)=\sum_{i=1}^M \left( \left[\frac{G_{fit}^{'}}{G_{exp}^{'}}-1\right]^2 + \left[\frac{G_{fit}^{''}}{G_{exp}^{''}}-1\right]^2 \right)
\label{eqn:costfunction}
\end{equation}
where M is the number of strain amplitude values measured for each mucus condition.

The steps of the fitting algorithm can be summarised as follows:
\begin{enumerate}
  \item Initial $G^*$, $n$ and $K^*$ are assumed ($\tau_y^*$ is determined beforehand, as explained above).
  \item At each strain amplitude, the resulting stress response is analysed by using Fourier transform rheology.
  \item $G^{'}$ and $G^{''}$ are obtained for the assumed model parameters and the given strain amplitude.
  \item The cost value is calculated from (\ref{eqn:costfunction}).
  \item If the cost function reaches its local minimum, then the algorithm stops and $G^*$, $n$ and $K^*$ are obtained. Otherwise, the values at the first step are modified and the algorithm runs until a local minimum is reached. To solve this minimisation problem Matlab's \citep{MATLAB:2020} "fmincon" function is used.
\end{enumerate}

\begingroup
\begin{table}
\centering
\def~{\hphantom{0}}
  \begin{tabular}{ccccc}
                                      & \bf{Healthy}  & \bf{Asthma}         & \bf{COPD}           & \bf{CF}    \\[3pt]
                                          \hline
\boldmath{$G^*$} \bf{(Pa)}           & 0.094          & 0.105          & 1.155          & 2.109 \\
\boldmath{$\tau_y^*$} \bf{(Pa)}      & 0.476          & 1.186          & 11.77$^a$      & 6.830 \\
\boldmath{$n$} \bf{(-)}              & 0.552          & 0.331          & 0.752          & 0.513 \\
\boldmath{$K^*$} ($\rm Pa\cdot s^n$) & 0.124          & 1.037          & 1.056          & 1.701 \\
\boldmath{$G_{\rm La=100}$} \bf{(-)} & \num{2.350e-3} & \num{2.625e-3} & \num{2.887e-2} & \num{5.272e-2} \\
\boldmath{$G_{\rm La=200}$} \bf{(-)} & \num{1.175e-3} & \num{1.312e-3} & \num{1.444e-2} & \num{2.636e-2} \\
\boldmath{$G_{\rm La=300}$} \bf{(-)} & \num{7.833e-4} & \num{8.750e-4} & \num{9.625e-3} & \num{1.757e-2} \\
\boldmath{$Wi_{\rm La=100}$} \bf{(-)}& 111.1          & 141.0          & 383.8          & 49.66 \\
\boldmath{$Wi_{\rm La=200}$} \bf{(-)}& 162.9          & 177.3          & 646.4          & 70.87 \\
\boldmath{$Wi_{\rm La=300}$} \bf{(-)}& 203.8          & 202.8          & 876.9          & 87.26 \\
\boldmath{$Bi_{\rm La=100}=\tau_{\rm y, La=100}$} \bf{(-)}& \num{1.189e-2} & \num{2.965e-2} & \num{2.942e-1} & \num{1.707e-1} \\
\boldmath{$Bi_{\rm La=200}=\tau_{\rm y, La=200}$} \bf{(-)}& \num{5.950e-3} & \num{1.482e-2} & \num{1.471e-1} & \num{8.537e-2} \\
\boldmath{$Bi_{\rm La=300}=\tau_{\rm y, La=300}$} \bf{(-)}& \num{3.967e-3} & \num{9.883e-3} & \num{9.808e-2} & \num{5.692e-2} \\
  \end{tabular}
\caption{The parameters of the Saramito-HB model determined by performing a rheological fitting procedure to the experimental results of \citet{patarin2020rheological}.\\
\scriptsize{$^a$ Taken from the Table 2 of \citet{patarin2020rheological} ($\sigma_c$ for COPD, spontaneous case)}}
\label{tab:rheologicalFitting}
\end{table}
\endgroup

The results of the fitting processes for the four conditions (healthy, asthma, COPD and CF) are given in table \ref{tab:rheologicalFitting}. It is clearly seen that the CF mucus displays the most elastic characteristics in contrast with the healthy mucus. The experiments on mucus simulants \citep{lafforgue2018rheological} and real mucus collected from subjects \citep{Hill2014} have shown that the mucus yield stress increases in the pathological conditions. The present results are consistent with this finding, and they are in line with the critical stresses presented in \citet{patarin2020rheological}.

\begin{figure}
  \centerline{\includegraphics[trim={0.0cm 0.0cm 0.0cm 0.0cm},clip,scale=0.9]{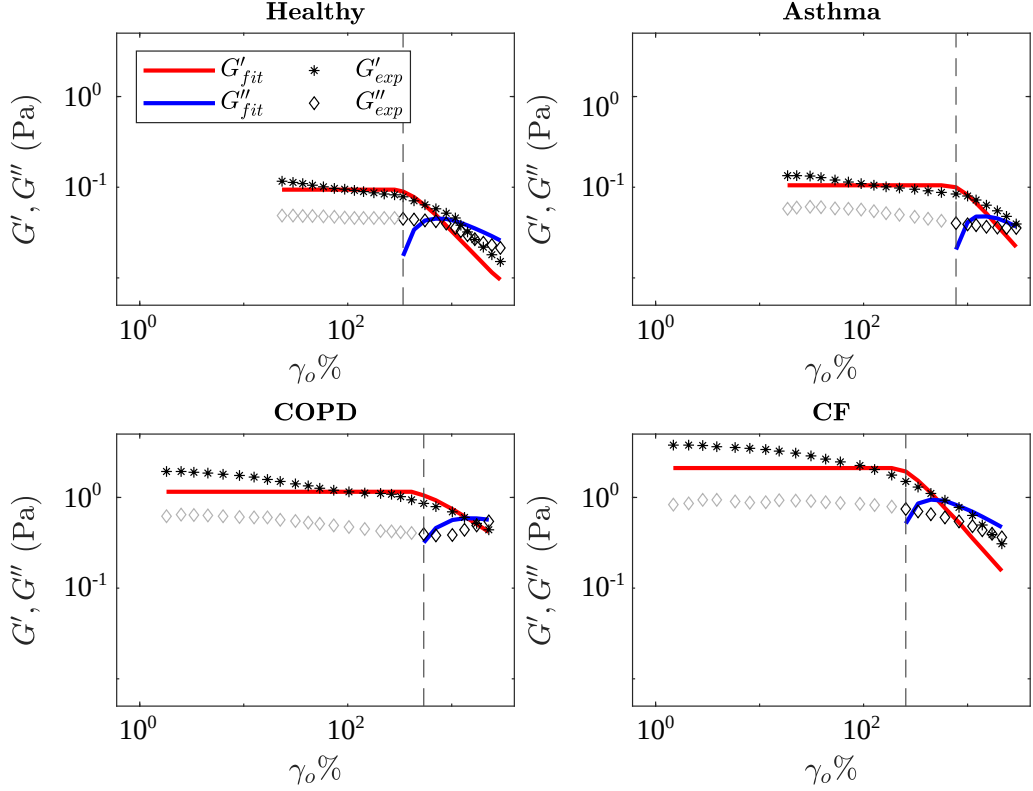}}
  \caption{ The viscoelastic moduli, $G^{'}$ and $G^{''}$ vs. strain amplitude, $\gamma_o\%$ for asthma, healthy, COPD, and CF mucus. The solid lines represent the viscoelastic moduli obtained after performing a parameter fitting to the experimental data of \citet{patarin2020rheological}, whose results are denoted by symbols. The vertical dashed black line represents the yield strain amplitude for each case. }
\label{fig:patarin20}
\end{figure}

The resulting $G^{'}$ and $G^{''}$ plotted against the strain amplitude, $\gamma_o\%$, are depicted in fig.\ \ref{fig:patarin20}. As seen from the vanishing $G^{''}$ and constant $G^{'}$, the Saramito-HB model predicts an ideal solid response until the material yields. This problem could be alleviated by using a kinematic hardening model, which predicts non-zero $G^{''}$ prior to yielding \citep{Fraggedakis2016sphericalparticle}. However, as it will be shown in the following sections, mucus adherent to the wall already yields just before and after the closure, when stress peaks occur, except extremely high-yield-stress cases (e.g. $\tau_y=2.5$). Therefore, the kinematic hardening concept is not included in the EVP model in the present study, and this is the reason why the markers for $G^{''}$ are grayed out in the unyielded region in figure \ref{fig:patarin20}. Furthermore, the experimental results of airway mucus rheology are usually prone to some problems such as saliva contamination \citep{joyner2019reliably}, effect of hypertonic saline solution (HSS) induction \citep{patarin2020rheological}, and small quantities of samples that can be obtained \citep{lock2018mucus}. These should be remembered when analysing such experimental studies on airway mucus samples.

%%%%%%%%%%%%%%%%%%%%%%%%%%%%%%%%%%%%%%%%%%%%%%%%%%%%%%%%%%
\subsection{Yielded zones for the healthy and the pathological conditions}\label{subsec:yieldingZones}

Table \ref{tab:rheologicalFitting} shows that healthy, asthma, COPD and CF mucus have different EVP characteristics. $G^{*}$ and $\tau_y^*$ of COPD and CF mucus samples are almost an order of magnitude larger than those of the healthy and asthma cases. Therefore in this section, the yielded zones are compared during an airway closure process of these cases with the parameters obtained in section \ref{subsec:rheologicalFitting}. In all cases, $\epsilon=0.35$ and $La=300$ are chosen to induce airway closure for COPD and CF cases, where otherwise, strong EVP characteristics inhibit the growth of the capillary instability that leads to airway closure. To determine the unyielded regions, a similar logic as \citet{chaparian2020yield} is used, so a criterion of $\textnormal{max}\left[ 0,\frac{\mid{\boldsymbol{\tau}^*}^d\mid - \tau_y^*}{K^*\mid{\boldsymbol{\tau}^*}^d\mid^n} \right] < 10^{-3}$ is set. The threshold of $10^{-3}$ is determined so that going lower than that does not alter the results substantially, but it prevents wiggles, which would be caused by the numerical nature of the study. 

\begin{figure}
  \centerline{\includegraphics[trim={0.0cm 0.0cm 0.0cm 0.0cm},clip,scale=0.9]{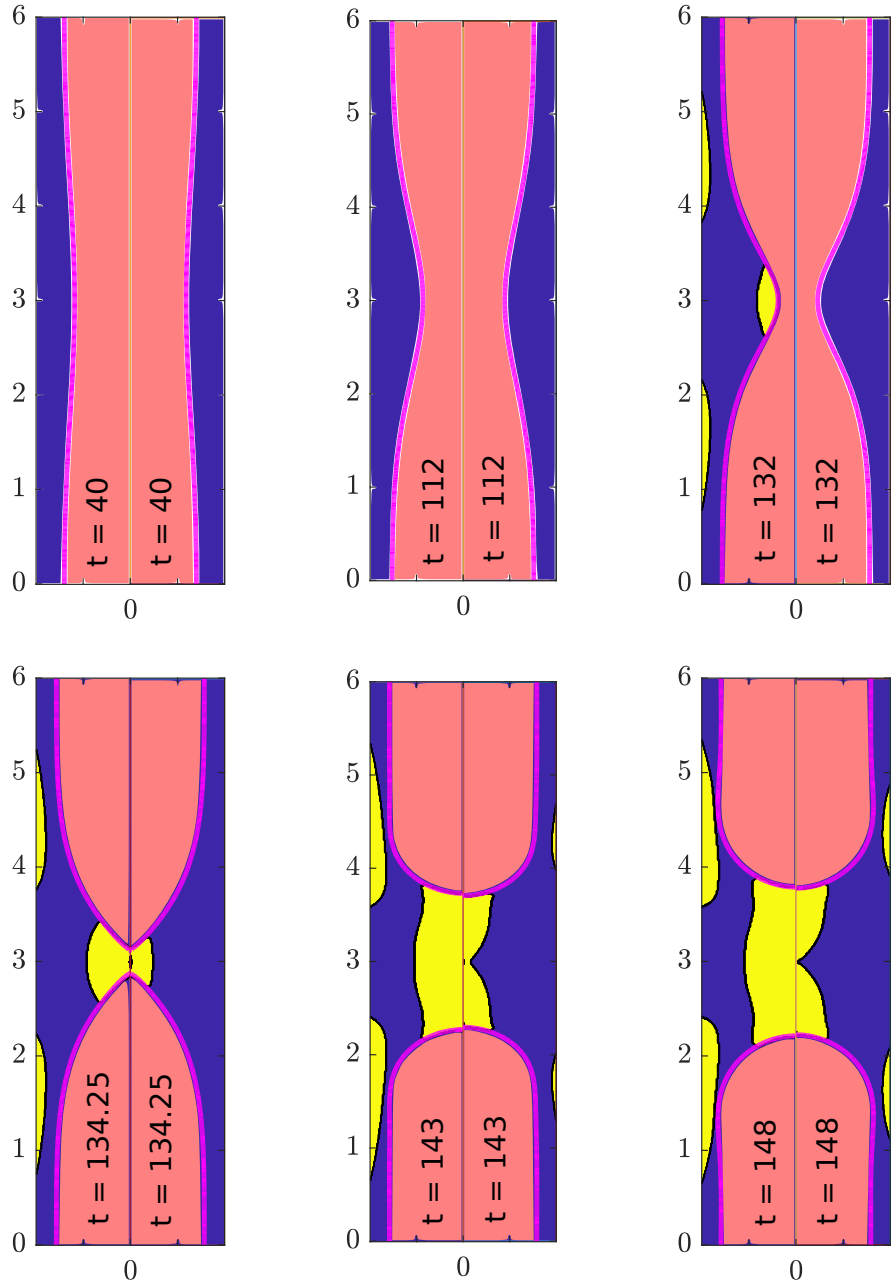}}
  \caption{ Comparison of yield zones between healthy (left-hand side of each panel) and asthma (right-hand side of each panel). Air is represented by light red. Yielded and unyielded regions of the liquid layer represented by yellow and blue, respectively. For EVP parameters of the liquid layer please refer to table \ref{tab:rheologicalFitting}. [$\lambda=6$, $\epsilon=0.35$ and $La=300$]. }
\label{fig:yieldZones_healthyVsAsthma}
\end{figure}

\begin{figure}
  \centerline{\includegraphics[trim={0.0cm 0.0cm 0.0cm 0.0cm},clip,scale=0.9]{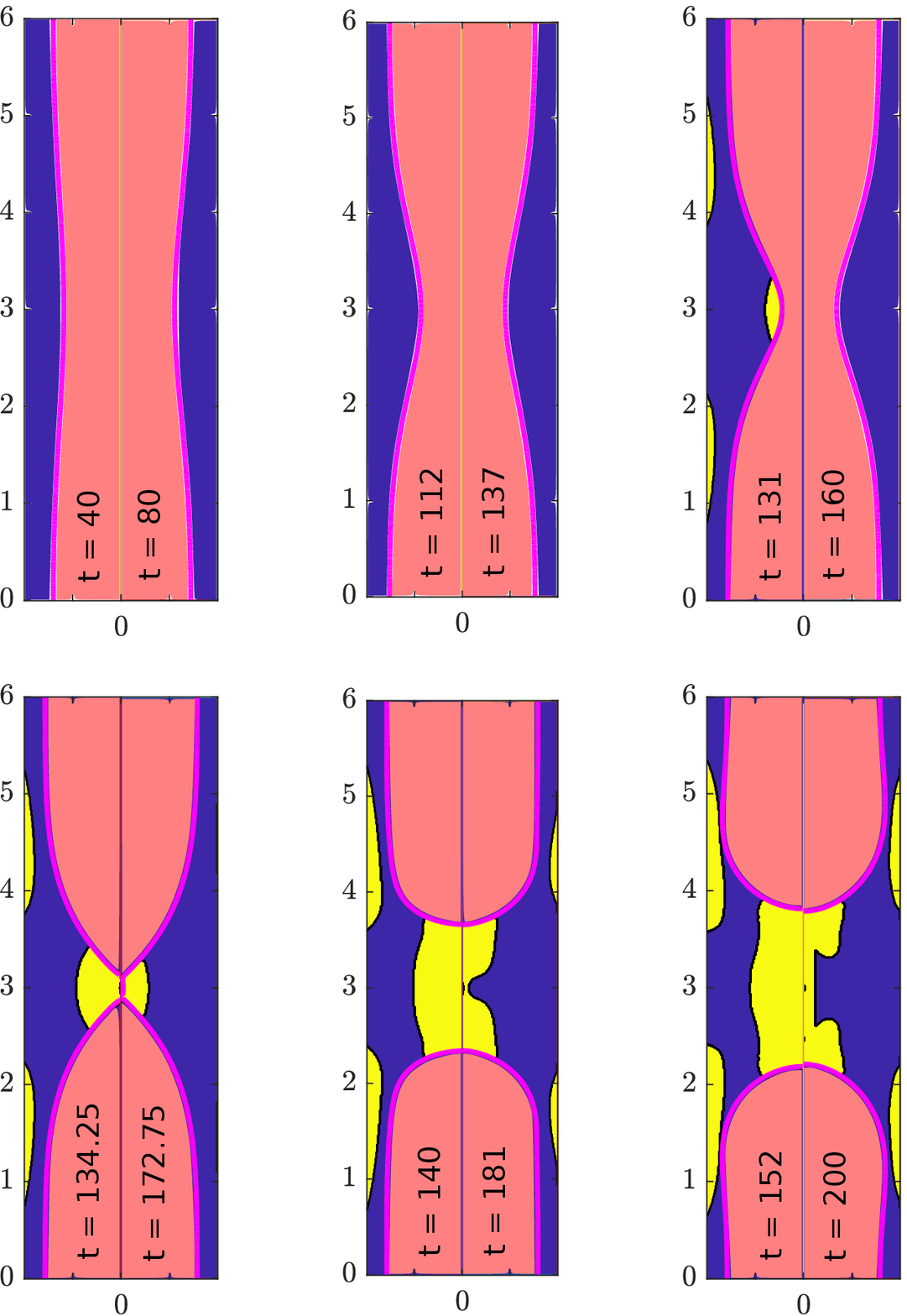}}
  \caption{ Comparison of yield zones between healthy (left-hand side of each panel) and COPD (right-hand side of each panel) mucus. Air is represented by light red. Yielded and unyielded regions of the liquid layer re represented by yellow and blue, respectively. For EVP parameters of the liquid layer please refer to table \ref{tab:rheologicalFitting}. [$\lambda=6$, $\epsilon=0.35$ and $La=300$]. }
\label{fig:yieldZones_healthyVsCopd}
\end{figure}

\begin{figure}
  \centerline{\includegraphics[trim={0.0cm 0.0cm 0.0cm 0.0cm},clip,scale=0.9]{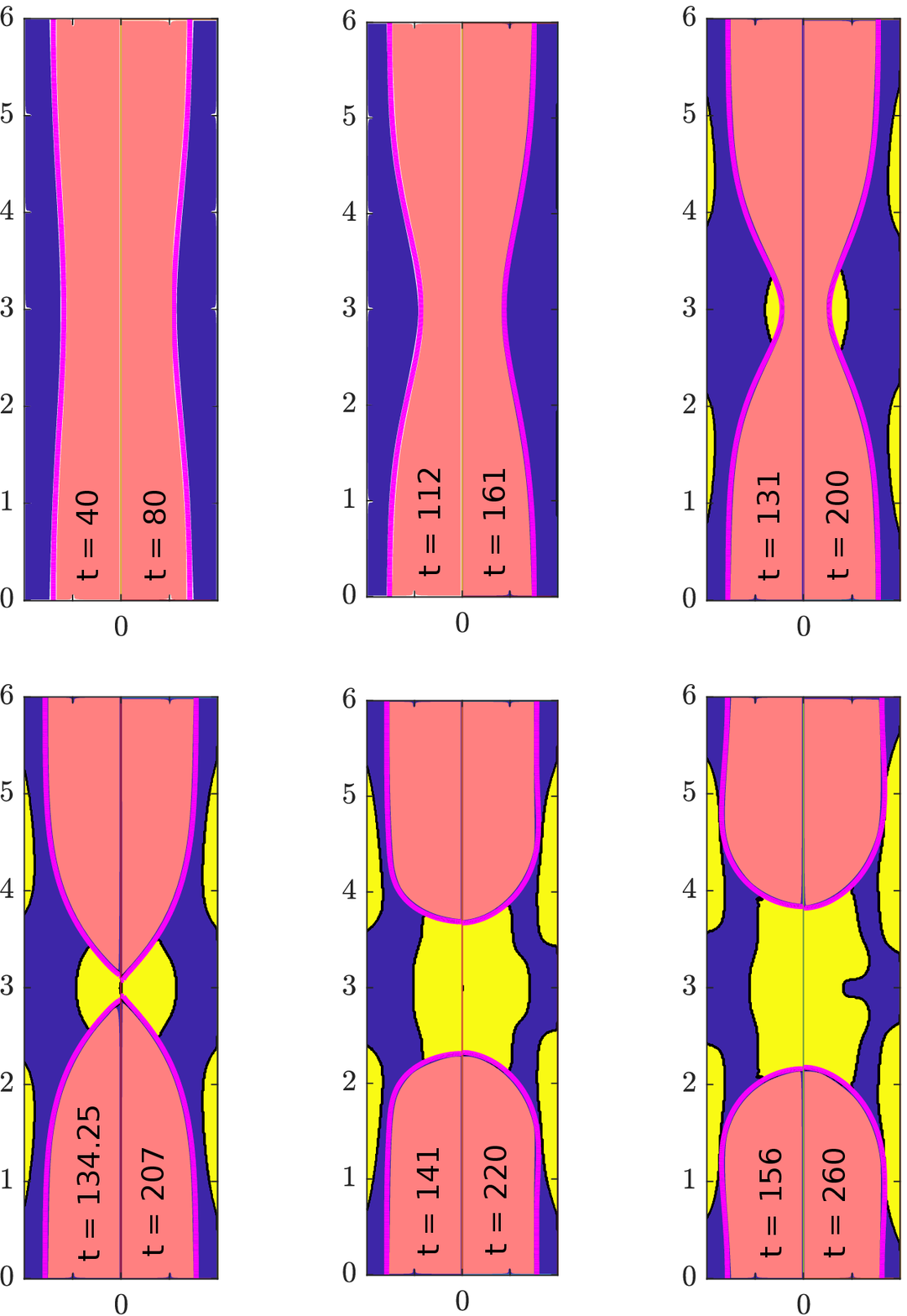}}
  \caption{ Comparison of yield zones between healthy (left-hand side of each panel) and CF (right-hand side of each panel) mucus. Air is represented by light red. Yielded and unyielded regions of the liquid layer re represented by yellow and blue, respectively. For EVP parameters of the liquid layer please refer to table \ref{tab:rheologicalFitting}. [$\lambda=6$, $\epsilon=0.35$ and $La=300$]. }
\label{fig:yieldZones_healthyVsCf}
\end{figure}

Figures \ref{fig:yieldZones_healthyVsAsthma}, \ref{fig:yieldZones_healthyVsCopd}, and \ref{fig:yieldZones_healthyVsCf} show how yielded and unyielded zones evolve with airway closure. In each figure, the left-hand side is the healthy case, and the right-hand sides are asthma, CF, and COPD cases, respectively. Newtonian air is represented by light red, and yielded and unyielded zones in EVP liquid layer are represented by yellow and blue, respectively. Snapshots are taken to show how yielded zones evolve with the deformation of the air-liquid interface, so there are three snapshots before and three after the closure event for each comparison.

Firstly, the healthy and asthma mucus are compared in figure \ref{fig:yieldZones_healthyVsAsthma}. As it was shown, EVP characteristics of these conditions are closer compared to that of the other two cases. The elastic modulus of the asthma mucus is 11.7\% more than that of healthy case, and its yield stress is more than double. Here, having a similar $G$ enables them to have almost the same nondimensional closure time ($t_c$), but due to its higher $\tau_y$, much less mucus yields during the plug formation in the asthma mucus. The EVP liquid layer is mainly in unyielded state especially before the breakup, so the bulk fluid is mainly affected by the elastic behaviour and solvent viscosity. Although it could not have been included in this paper due to brevity, this has been confirmed in a parametric study, where $\tau_y^*$ and $G^*$ of the material has been individually varied in our parameter range, and it has been found that $G^*$ is the most significant parameter responsible for the closure time. This study has showed that even for the maximum value of yield stress in our parameter range, the closure occurred with a negligible delay, although there was no yielded regions. The liquid layer was able to bend as an elastic solid when $G$ was not significantly large. Moreover, in both the healthy and asthma cases, yielded zones concentrate at the plug tip and near the wall close to the shoulder, where shear stresses prevail.

Then, the healthy and COPD mucus cases are compared in figure \ref{fig:yieldZones_healthyVsCopd}. The rheological fitting in section \ref{subsec:rheologicalFitting} resulted in an order of magnitude larger $\tau_y$ and $G$ for COPD mucus than healthy one. Larger $G$ of the COPD mucus results in a $\Delta{t_c}=38.5$ difference between the nondimensional closure times of these cases due to increased stiffness of the mucus. This finding is consistent with the previous interpretation from \citet{Romano2021}, where it was shown that $Wi$ has a significant impact on the closure time (note that $Wi$ is inversely proportional to $G$). Furthermore, the COPD mucus has smaller yielded zone due to its higher $\tau_y$, and these are located at the bulge tip and just behind the shoulder as in the previous case, since shear stresses are larger in these regions.

Finally, figure \ref{fig:yieldZones_healthyVsCf} compares the yielded zones of healthy and CF mucus during the closure. Here, the CF mucus behaves somewhat counterintuitively compared to the other cases because despite its larger $\tau_y$, it has a larger yielded zone compared to the healthy case. However, it should be noted that its $G$ is also 82.6\% larger than that of the COPD case, so actually, the COPD and CF cases have fairly different Weissenberg numbers ($\rm Wi_{\rm CF}=87.26$ and $Wi_{\rm COPD}=876.9$, see table \ref{tab:rheologicalFitting}). \citet{izbassarov2020dynamics} presented a complete yielding regime map for an EVP droplet in a Newtonian medium in a certain $Wi$ and $Bi$ interval. They showed that when $Wi$ of the droplet decreases ($G$ increases), it can yield even at higher $Bi$. Therefore, it would be interesting to further study the interplay between $Wi$ and $Bi$ in yielded regions in this interfacial instability problem.

%%%%%%%%%%%%%%%%%%%%%%%%%%%%%%%%%%%%%%%%%%%%%%%%%%%%%%%%%%
\subsection{Effect of pathological conditions}\label{subsec:effectOfPathological}

The healthy, asthma, COPD, CF and Newtonian conditions are compared in terms of wall shear stress and pressure excursions and their local gradients. The EVP parameters obtained in section \ref{subsec:rheologicalFitting} are used to simulate the healthy, asthma, COPD and CF mucus. Newtonian case is also simulated just to compare how the EVP characteristics of the liquid layer affect the mechanical stresses. Furthermore, $\epsilon$ and $La$ are varied to study conditions, such as mucus hypersecretion and surfactant deficiency, and also to induce the closure for highly viscoplastic cases. The results of this section are presented in the following three subsections, where $La=100$, $La=200$, and $La=300$, respectively.

\subsubsection{$La=100$}\label{subsubsec:La100} %%*******************%%
\begin{figure}
  \centerline{\includegraphics[trim={0.0cm 0.0cm 0.0cm 0.0cm},clip,scale=0.9]{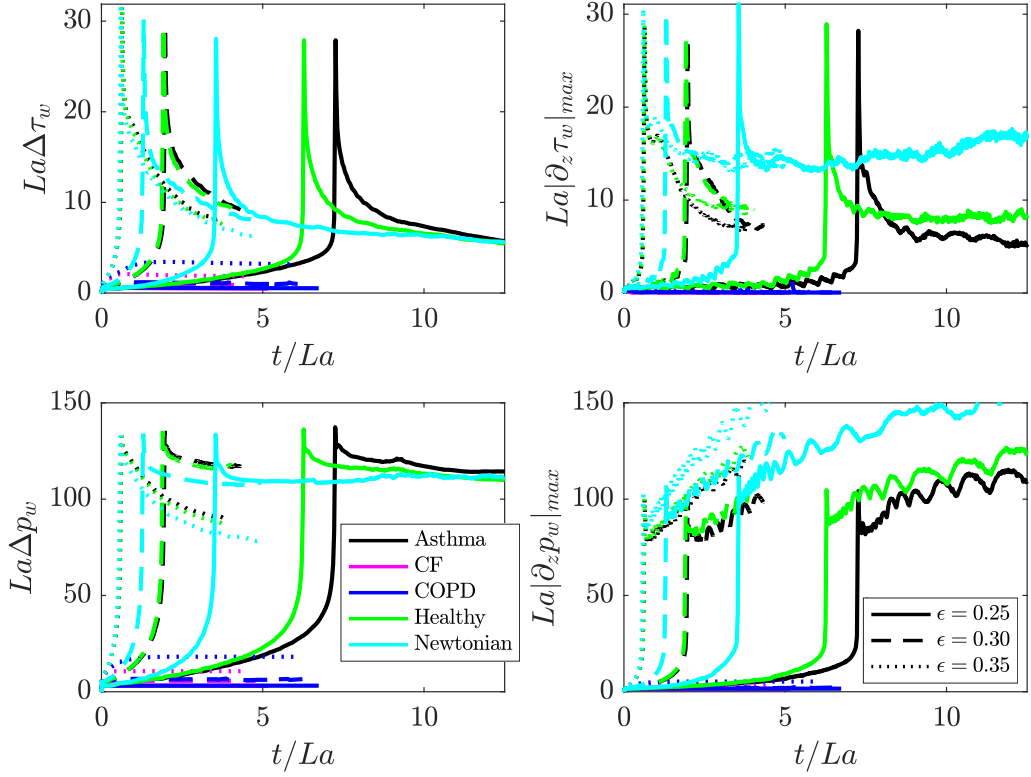}}
  \caption{ The effects of pathological conditions for $La=100$. $\epsilon$ is also varied in each panel. Time evolutions of the wall shear stress excursions $\Delta \tau_w =$ max($\tau_w$) - min($\tau_w$) (top left), the maximum absolute value of the wall shear stress gradient $|\partial_z \tau_w|_{\rm max}$ (top right), the wall pressure excursions $\Delta p_w =$ max($p_w$) - min($p_w$) (bottom left), and the maximum absolute value of the wall pressure gradient $|\partial_z p_w|_{\rm max}$ (bottom right) for asthma, CF, COPD, healthy, and Newtonian cases for $\epsilon=0.25$, $\epsilon=0.30$, and $\epsilon=0.35$. Note that the non-dimensional time, $t$, is divided by $La$ to eliminate the effect of surface tension, $\sigma^*$, on the time scaling for a better interpretation of the results. $\Delta \tau_w$, $|\partial_z \tau_w|_{\rm max}$, $\Delta p_w$, and $|\partial_z p_w|_{\rm max}$ are also re-scaled by $La$ for the same purpose. }
\label{fig:pathologicalLa100}
\end{figure}

The conditions in this section describe a part of an airway with relatively normal surfactant activity with increasingly severe mucus hypersecretion for asthma, CF, COPD, healthy and Newtonian liquid layers. In figure \ref{fig:pathologicalLa100}, wall tangential and normal stress local gradients, as well as the stress excursions are depicted for $La=100$ for $\epsilon=0.25$, $\epsilon=0.30$, and $\epsilon=0.35$. Mechanical stresses and non-dimensional time ($t$) are re-scaled by $La$ to eliminate the effect of surface tension in the original scaling and to better interpret the results. It is seen that strong EVP features of CF and COPD mucus inhibit airway closure at this $La$ value. The maximum values of the stress peaks for the healthy and asthma cases are almost the same regardless of $\epsilon$, since these initial stress peaks are mostly related to the Newtonian nature of the liquid \citep{Romano2021}. We moreover note that $t_c$ moves closer and converges to the Newtonian case in both healthy and asthma cases as $\epsilon$ increases indicating that the effect of pathological conditions on the closure time diminishes as the initial liquid layer thickness increases.

\begin{figure}
  \centerline{\includegraphics[trim={0.0cm 0.0cm 0.0cm 0.0cm},clip,scale=0.90]{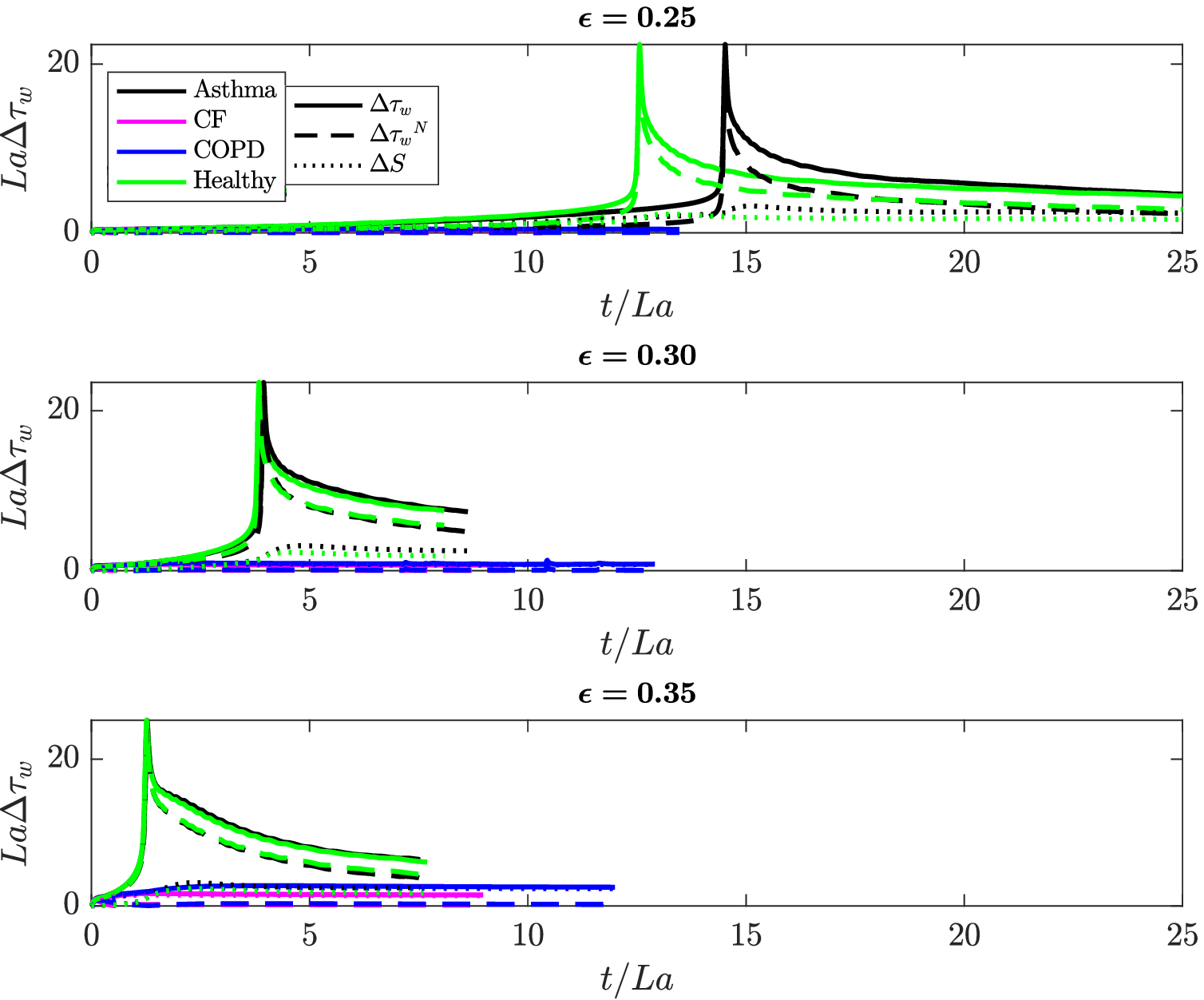}}
  \caption{ Evolution of wall tangential stress excursion, $\Delta \tau_w =$ max($\tau_w$) - min($\tau_w$), which consists of Newtonian ($\Delta{\tau}^N$) and extra-stress ($\Delta{S}$) components, for $La=100$ and  $\epsilon=0.25$ (top), $\epsilon=0.30$ (middle), and $\epsilon=0.35$ (bottom). Note that the non-dimensional time, $t$, is divided by $La$ to eliminate the effect of surface tension, $\sigma^*$, on the time scaling for a better interpretation of the results. $\Delta \tau_w$ is also re-scaled by $La$ for the same purpose. }
\label{fig:pathologicalLa100Extra}
\end{figure}

Shear stress excursion, $\Delta \tau_w$, is decomposed into its extra-stress, $\Delta{S}$, and Newtonian, $\Delta{\tau}^N$, components in figure \ref{fig:pathologicalLa100Extra}. As it was discussed earlier, the healthy and asthma mucus have weak viscoelastic and viscoplastic characteristics, so except for a minor increase in $\Delta{S}$ after the closure, there is no significant effect of them on the post-coalescence dynamics. Also, initial peak is almost solely due to the Newtonian component, as it was also pointed out by \citet{Romano2021}.

\subsubsection{$La=200$}\label{subsubsec:La200} %%*******************%%

\begin{figure}
  \centerline{\includegraphics[trim={0.0cm 0.0cm 0.0cm 0.0cm},clip,scale=0.9]{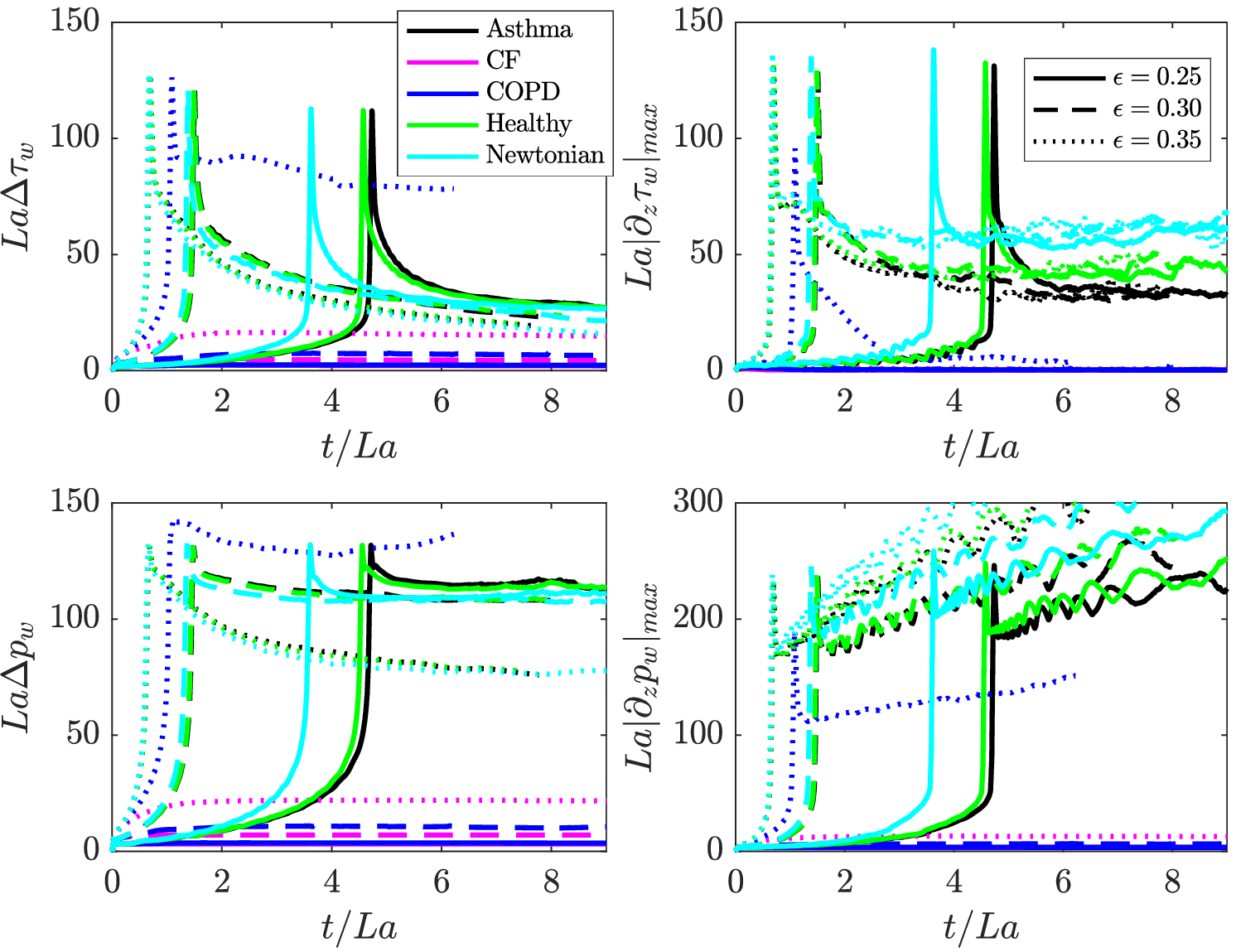}}
  \caption{ The effects of pathological conditions for $La=200$. $\epsilon$ is also varied in each panel. Time evolutions of the wall shear stress excursions $\Delta \tau_w =$ max($\tau_w$) - min($\tau_w$) (top left), the maximum absolute value of the wall shear stress gradient $|\partial_z \tau_w|_{\rm max}$ (top right), the wall pressure excursions $\Delta p_w =$ max($p_w$) - min($p_w$) (bottom left), and the maximum absolute value of the wall pressure gradient $|\partial_z p_w|_{\rm max}$ (bottom right) for asthma, CF, COPD, healthy, and Newtonian cases for $\epsilon=0.25$, $\epsilon=0.30$, and $\epsilon=0.35$. Note that the non-dimensional time, $t$, is divided by $La$ to eliminate the effect of surface tension, $\sigma^*$, on the time scaling for a better interpretation of the results. $\Delta \tau_w$, $|\partial_z \tau_w|_{\rm max}$, $\Delta p_w$, and $|\partial_z p_w|_{\rm max}$ are also re-scaled by $La$ for the same purpose. }
\label{fig:pathologicalLa200}
\end{figure}

After studying the airway closure in $La=100$, the surface tension of the air-liquid interface is increased, and the system is analysed when $La=200$. The results are presented in the same fashion as in the previous subsection. Figure \ref{fig:pathologicalLa200} shows that closure times for the healthy, asthma and Newtonian cases are short, and the stresses are higher compared to the $La=100$ case as expected. However, the biggest difference is that there is a liquid plug formation for the COPD mucus at this condition, but its closure is slower compared to the healthy and asthma cases due to its higher $G$. Its tangential and normal stress excursion peaks are around the same compared to the other cases, where closure occurs. However, the local gradients of these stresses are significantly lower. This will be discussed in subsection \ref{subsubsec:detailedAnalysisOfTheLower}.

\begin{figure}
  \centerline{\includegraphics[trim={0.0cm 0.0cm 0.0cm 0.0cm},clip,scale=0.90]{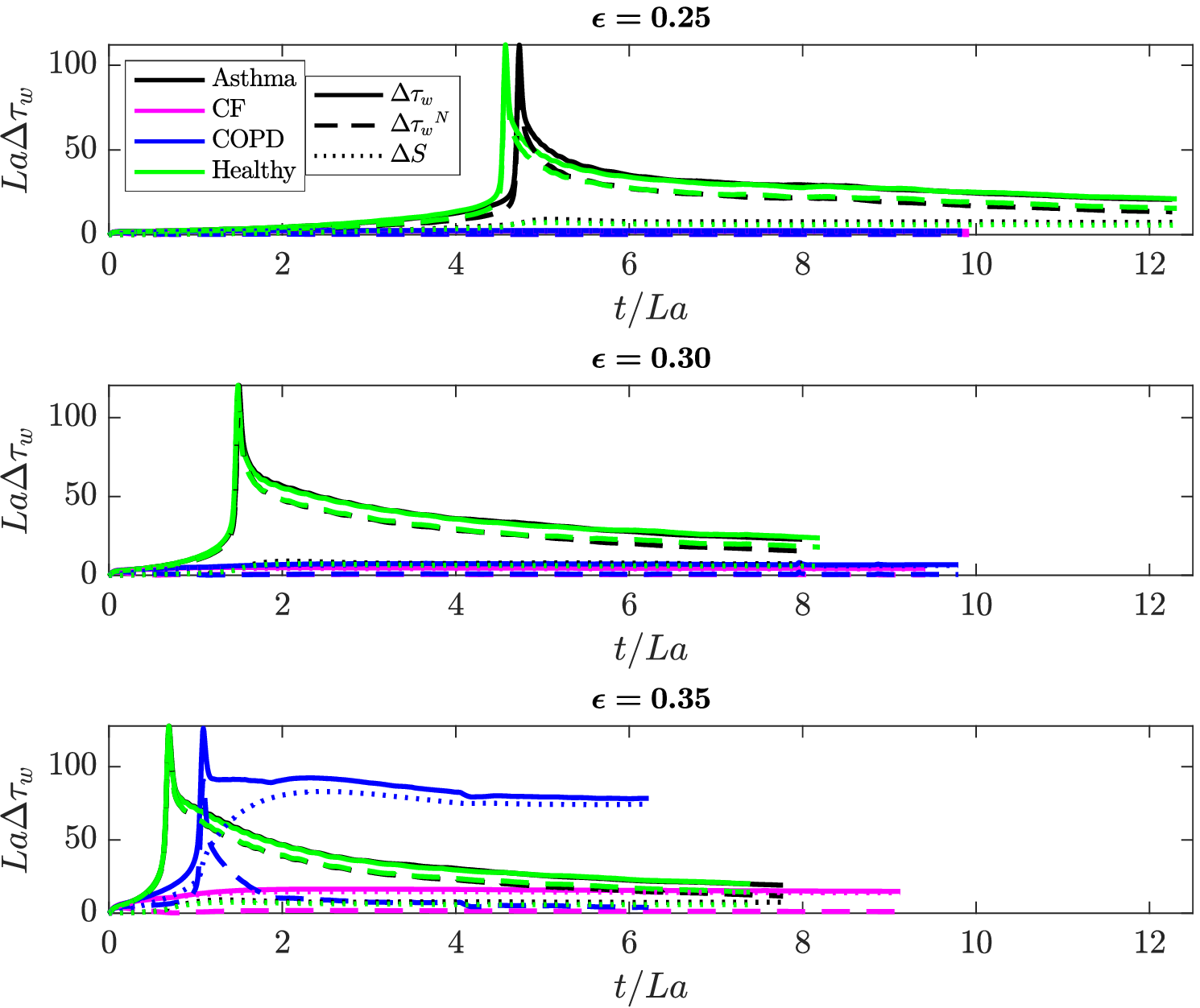}}
  \caption{ Evolution of wall tangential stress excursion, $\Delta \tau_w =$ max($\tau_w$) - min($\tau_w$), which consists of Newtonian ($\Delta{\tau}^N$) and extra-stress ($\Delta{S}$) components, for $La=200$ and  $\epsilon=0.25$ (top), $\epsilon=0.30$ (middle), and $\epsilon=0.35$ (bottom). Note that the non-dimensional time, $t$, is divided by $La$ to eliminate the effect of surface tension, $\sigma^*$, on the time scaling for a better interpretation of the results. $\Delta \tau_w$ is also re-scaled by $La$ for the same purpose. }
\label{fig:pathologicalLa200Extra}
\end{figure}

The tangential wall stress excursion is decomposed into its Newtonian and extra stress components for this case as well as in fig.\ \ref{fig:pathologicalLa200Extra}. For the less EVP cases (healthy and asthma), the contribution of extra stress to the total $\Delta \tau_w$ is very low as in $La=100$. However, for the COPD mucus the initial peak increases almost by 30\% due to the increase in the extra stress contribution to the total tangential stress excursion on the wall. This indicates that in highly EVP mucus, the peak of the stresses may not be solely due to the Newtonian contribution, but also to the extra stress contribution. The extra stress keeps growing as the Newtonian component relaxes after the closure. Another major point for COPD closure is that extra stress persists after the closure, and its magnitude is around the same levels as the Newtonian peak. This suggests that highly EVP mucus damages the cells on the respiratory wall continuously, which is opposite to the Newtonian case, where the stresses relax to lower levels after reaching their peak values.

\subsubsection{$La=300$}\label{subsubsec:La300} %%*******************%%

\begin{figure}
  \centerline{\includegraphics[trim={0.0cm 0.0cm 0.0cm 0.0cm},clip,scale=0.9]{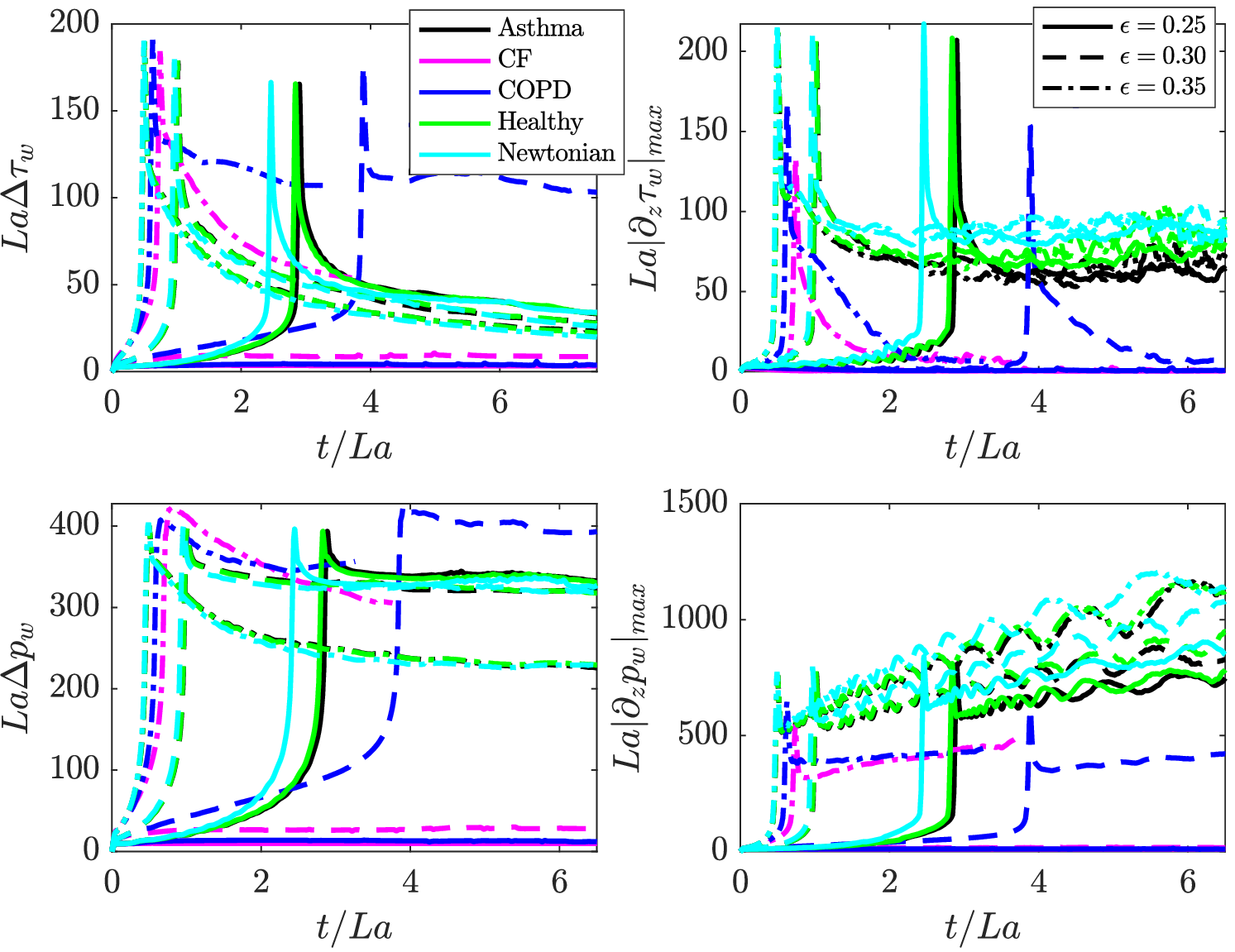}}
  \caption{ The effects of pathological conditions for $La=300$. $\epsilon$ is also varied in each panel. Time evolutions of the wall shear stress excursions $\Delta \tau_w =$ max($\tau_w$) - min($\tau_w$) (top left), the maximum absolute value of the wall shear stress gradient $|\partial_z \tau_w|_{\rm max}$ (top right), the wall pressure excursions $\Delta p_w =$ max($p_w$) - min($p_w$) (bottom left), and the maximum absolute value of the wall pressure gradient $|\partial_z p_w|_{\rm max}$ (bottom right) for asthma, CF, COPD, healthy, and Newtonian cases for $\epsilon=0.25$, $\epsilon=0.30$, and $\epsilon=0.35$. Note that the non-dimensional time, $t$, is divided by $La$ to eliminate the effect of surface tension, $\sigma^*$, on the time scaling for a better interpretation of the results. $\Delta \tau_w$, $|\partial_z \tau_w|_{\rm max}$, $\Delta p_w$, and $|\partial_z p_w|_{\rm max}$ are also re-scaled by $La$ for the same purpose. }
\label{fig:pathologicalLa300}
\end{figure}

To investigate the closure for surfactant-deficient conditions, $\sigma^*$ is increased further, and the results are presented in figure \ref{fig:pathologicalLa300}. The trends for $t_c$ and the peaks of stresses seen in $La=200$ case continues for $La=300$ as well. However, due to the extreme conditions for $La$ and $\epsilon$, the COPD mucus forms a plug at $\epsilon=0.30$ and $\epsilon=0.35$, and the CF mucus also forms a plug at $\epsilon=0.35$. The closure time $t_c$ of the CF mucus is slightly delayed, but its peaks for $\Delta \tau_w$ and $\Delta p_w$ are around the same levels compared to the other cases. Furthermore, tangential and normal local stress gradients are lower as it was seen in figure \ref{fig:pathologicalLa200} for the COPD mucus. 

\begin{figure}
  \centerline{\includegraphics[trim={0.0cm 0.0cm 0.0cm 0.0cm},clip,scale=0.90]{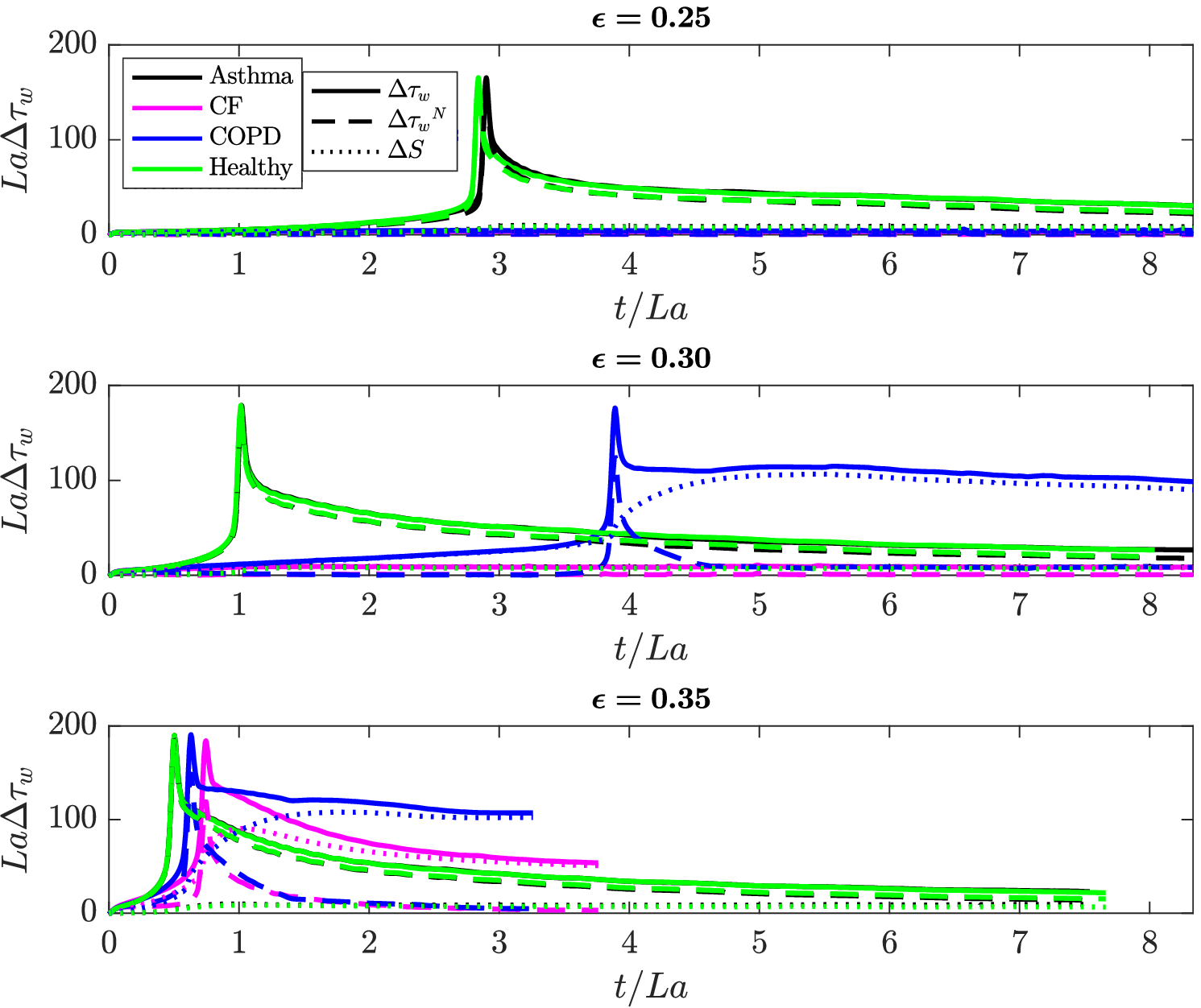}}
  \caption{ Evolution of wall tangential stress excursion, $\Delta \tau_w =$ max($\tau_w$) - min($\tau_w$), which consists of Newtonian ($\Delta{\tau}^N$) and extra-stress ($\Delta{S}$) components, for $La=300$ and  $\epsilon=0.25$ (top), $\epsilon=0.30$ (middle), and $\epsilon=0.35$ (bottom). Note that the non-dimensional time, $t$, is divided by $La$ to eliminate the effect of surface tension, $\sigma^*$, on the time scaling for a better interpretation of the results. $\Delta \tau_w$ is also re-scaled by $La$ for the same purpose. }
\label{fig:pathologicalLa300Extra}
\end{figure}

The extra stress contribution is also checked for the $La=300$ case in figure \ref{fig:pathologicalLa300Extra}. The results show similar characteristics to the $La=200$ case. However, there are two important points to note. The first one is that extra stress component almost doubles the initial peak for the CF mucus. After the peak, the stress relaxes very slowly compared to the healthy and asthma conditions. The second point is that the stresses in the CF mucus relaxes unlike the COPD mucus, where extra stress persists for the entire duration of the simulation. The COPD mucus has higher $\tau_y$ than that of the CF mucus, therefore it is interpreted that this makes the relaxation of the stresses more difficult, and as a result, the airway epithelial cells are exposed to high levels of stress excursions as long as the plug exists. This phenomenon has not been observed by the viscoelastic simulations of \citet{Romano2021}, where they reported a secondary peak of the stresses after relaxation from the initial Newtonian peak.

\subsubsection{Detailed analysis of the lower local stress gradients in the CF and COPD conditions}\label{subsubsec:detailedAnalysisOfTheLower} %%*******************%%

\begin{figure}
  \centerline{\includegraphics[trim={0.0cm 0.0cm 0.0cm 0.0cm},clip,scale=0.9]{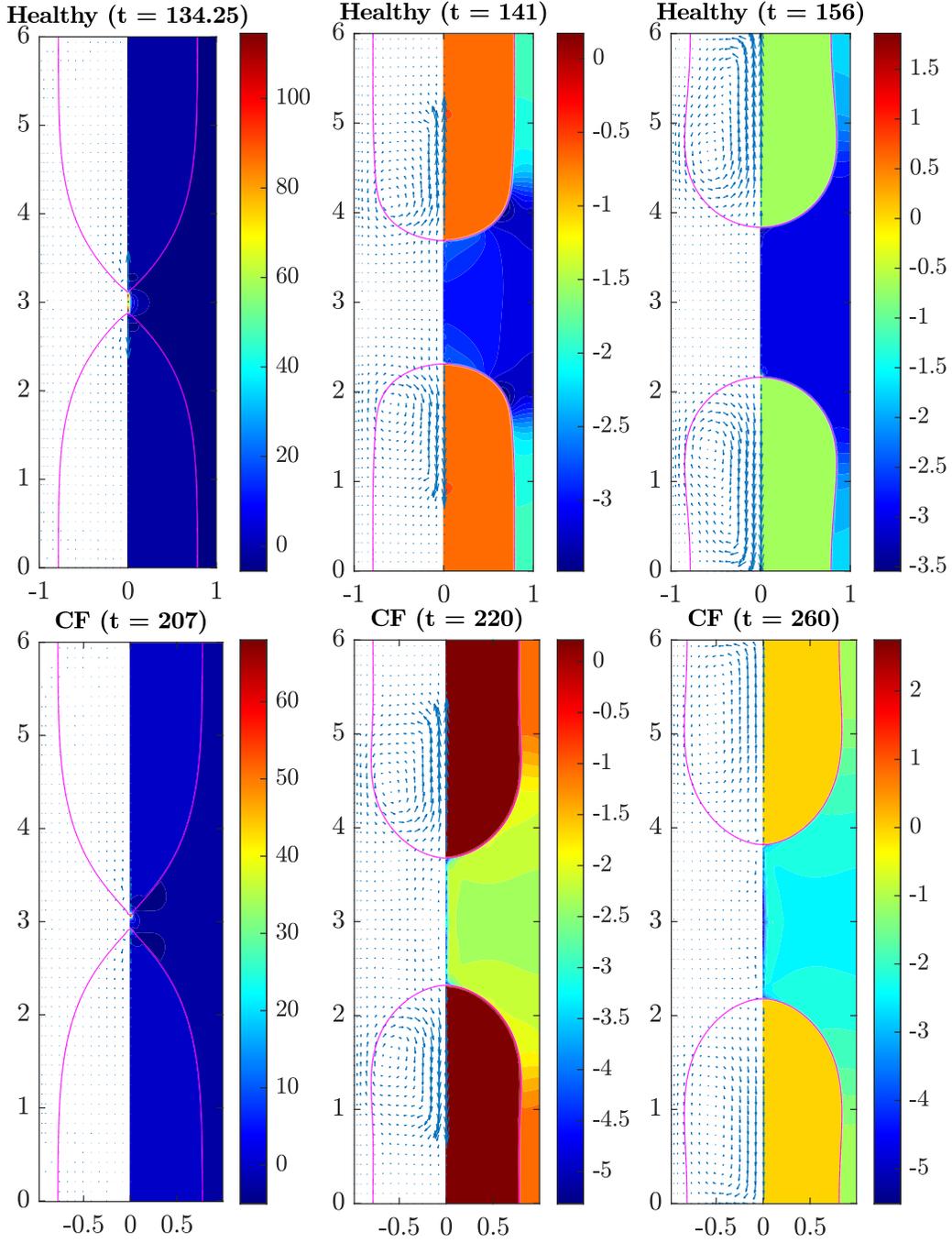}}
  \caption{ Evolution of the interfaces (solid magenta lines) with constant contours of the pressure ﬁeld (right portion of each subplot) and the velocity vectors (left portion of each subplot) for the healthy and the CF mucus. [$La=300$ and $\epsilon=0.35$]. }
\label{fig:presVelProfile_cfVShealthy}
\end{figure}

To further analyse the lower peak of $|\partial_z \tau_w|_{\rm max}$ and $|\partial_z p_w|_{\rm max}$ for the COPD and CF mucus cases observed in figures \ref{fig:pathologicalLa200} and \ref{fig:pathologicalLa300}, their pressure and velocity fields are plotted in figure \ref{fig:presVelProfile_cfVShealthy}. The pressure contours are plotted on the right-hand side of each panel, and the velocity vectors are plotted on the left-hand side. Also, the top and bottom rows represent the healthy and CF mucus cases, respectively. \citet{Romano2019} stated that the peak of the local normal stress gradient is related to the capillary wave formed after the plug formation. When the air-liquid interface is compared between the healthy and the CF cases, it can be seen that the curvature formed by the healthy case is larger than that of the CF case (especially top-right and bottom-right snapshots). The larger elastic modulus of CF mucus makes the liquid layer stiffer, so its interface has a lower curvature compared to that of the healthy one. This can also be confirmed by the higher pressure gradient around the shoulder of the healthy mucus compared to the CF case.

\begin{figure}
  \centerline{\includegraphics[trim={0.0cm 0.0cm 0.5cm 0.0cm},clip,scale=0.9]{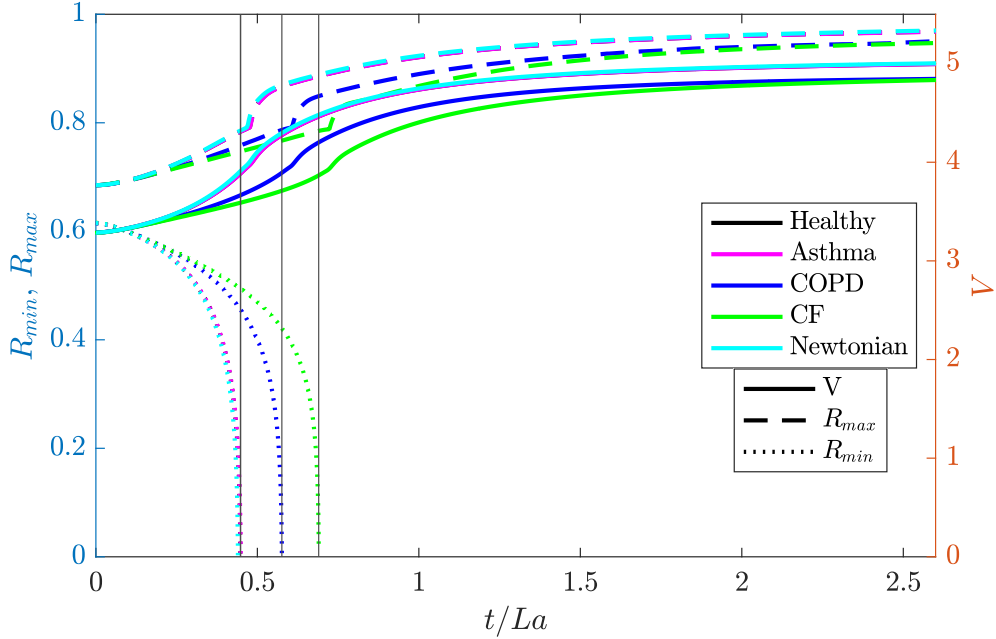}}
  \caption{ Time evolution of the minimum and maximum core radius of air-liquid interface ($R_{\rm min}$ and $R_{\rm max}$), and the mucus layer volume between the non-dimensional axial locations of $z=1.3$ and $z=4.7$ ($V$) for healthy, asthma, COPD, CF, and Newtonian cases. The closure times are denoted by a black vertical solid line for each case. Note that the non-dimensional time, $t$, is divided by $La$ to eliminate the effect of surface tension, $\sigma^*$, on the time scaling for a better interpretation of the results. [$La=300$ and $\epsilon=0.35$]. }
\label{fig:statPathologiesWithNewt}
\end{figure}

This analysis is quantified in figure \ref{fig:statPathologiesWithNewt} by plotting the time evolution of the minimum and maximum core radius of air-liquid interface ($R_{\rm min}$ and $R_{\rm max}$), and the mucus layer volume between the non-dimensional axial locations of $z=1.3$ and $z=4.7$ ($V$) for the healthy, asthma, COPD, CF, and Newtonian cases. It should be noted that where $R_{\rm max}$ is the highest, the liquid layer thickness is the smallest. The figure clearly shows that the maximum value of $R_{\rm max}$ is smaller for the COPD and CF mucus cases, so this confirms that the air-liquid interface does not bend easily, and forms a capillary wave with a smaller curvature in these cases. Hence, the result is lower local normal and tangential stress peaks. This is also in agreement with the asymptotic theory for a Bretherton bubble \cite[see][]{bretherton1961theMotion} as Roman\`o et al.\cite{romano2022theEffect} pointed out, i.e. $\partial_z{p_w} \approx -\epsilon \partial_z^3 (1-R_I)$.

Another important point is that the higher $G$ of the COPD and CF mucus cases slow down the liquid accumulation at the center, thus making the whole process slower. This slower rate of liquid transfer to the liquid bulge results in lower velocity gradient around the wall, and consequently, lower local shear stress gradients. A similar smearing effect has also been related to the increase of viscosity of the liquid layer before \citep{Romano2019, erken2022capillary}.

%%%%%%%%%%%%%%%%%%%%%%%%%%%%%%%%%%%%%%%%%%%%%
%%%%%%%%%%%%%%%%%%%%%%%%%%%%%%%%%%%%%%%%%%%%%
\section{Summary and conclusions}\label{sec:summaryAndConclusions}
The effects of the non-Newtonian characteristics of the mucus on airway closure have been studied in a model problem, where an elastoviscoplastic (EVP) liquid layer coats inside of a rigid pipe and surrounds the air core. The rheological properties of the EVP liquid layer have been determined by fitting the Saramito-HB model to the experimental data for the healthy, asthma, COPD and CF mucus cases \citep{patarin2020rheological}. These mucus conditions are studied in varying Laplace number and initial mucus thickness conditions, and the possible effects on the wall stresses are analysed.

Firstly, the EVP parameters for four different conditions of airway mucus have been obtained by rheological fitting process similar to Fraggedakis et al.\cite{Fraggedakis2016evpcomparison}. Here, the viscoelastic moduli $G^{'}$ and $G^{''}$ are fitted to the experimental data by \citet{patarin2020rheological}, and the EVP model parameters $G^*$, $n$, $K^*$, and $\tau_y^*$ are determined for the healthy, asthma, COPD and CF cases. These parameters are used to examine the effects of the pathological conditions. 

Yielded zones of asthma, COPD and CF mucus are compared against the healthy one in both pre- and post-coalescence phases. In this comparison, it is found that the slightly higher $\tau_y^*$ of the asthma mucus causes less yielding during the closure. However, the profile of the yielded zones are very similar in both cases, and concentrated at the bulge tip and at the wall near the shoulder, where the tangential stresses are larger. On the other hand, the COPD mucus is an order of magnitude larger $\tau_y^*$ and $G^*$ than those of the healthy mucus. The larger $G^*$ results in a significant delay of the closure, while the larger $\tau_y^*$ increases the quantity of unyielded mucus, and results in a more solid-like behavior.  Lastly, the comparison is made between the healthy and the CF conditions. Here, it is shown that the CF mucus has considerably larger yielded regions than that of the healthy one. This counterintuitive behaviour is attributed to the much larger elastic modulus of the CF mucus compared to the COPD mucus, which alters the Weissenberg number of both layers significantly. A similar behaviour is reported by \citet{izbassarov2020dynamics} for an EVP droplet in a Newtonian medium. In all cases, the EVP liquid layer is predominantly in unyielded state before the breakup suggesting that the elastic behaviour and solvent viscosity are the main factors affecting the bulk fluid behaviour before the closure.

Afterwards, the effect of the pathological conditions on the wall stresses are analysed by varying surface tension (Laplace number) and initial mucus thickness conditions as recent findings suggest that these can be observed in pathological conditions of the airway \citep{de2018airway,agudelo2020decreased, griese1997pulmonary}. This shows that the large $G$ (small $Wi$) of the COPD and CF mucus inhibits the plug formation at low $La$ values. It confirms that the elastic behaviour of the fluid is dominant before the breakup and mainly affects whether the closure occurs. The closure time, $t_c$, is also largely determined by $G$. The initial undisturbed liquid layer thickness, $\epsilon$, was also varied at each studied $La$ value to find out how the wall stresses and $t_c$ are affected by this parameter, and it is found that as $\epsilon$ increases $t_c$ converges to the Newtonian case in each pathological case. 

The influence of the pathological cases on the wall stresses are also studied as these are important to estimate how the airway epithelium may be affected in these conditions. It is seen that the contribution from the viscoelastic stress ($\Delta{S}$) on the total stress ($\Delta{\tau}$) at the peak stress levels is higher in the COPD and CF cases. Furthermore, $\Delta{S}$ relaxes very slowly after the closure, and stays almost as high as the Newtonian peak for a very long time. This continuous disturbance after the closure may increase the damage on the pulmonary epithelial cells. 

Another important finding is that the high nondimensional stiffness ($G$) of the COPD and CF mucus causes smaller curvature at the capillary wave, and consequently smaller peaks of local normal stress gradients during the closure. The local tangential stress gradients are also smaller for the COPD and CF conditions because of the slower accumulation of liquid at the axial center of the domain. This indicates that the main source of stress for the pulmonary epithelium may be coming from the tangential and normal stress excursions in the COPD and CF mucus ($\Delta{\tau_w}$ and $\Delta{p_w}$).

\begin{comment}
    \begin{itemize}
      \item Large $G$ of the COPD and CF mucus cases inhibits the plug formation at low $La$ values.
      \item As $\epsilon$ increases $t_c$ converges to the Newtonian case in each pathological case.
      \item Strong EVP characteristics of the COPD and CF mucus increases the contribution from the viscoelastic stress ($\Delta{S}$) on the total stress ($\Delta{\tau}$) at the peak stress levels.
      \item $\Delta{S}$ relaxes very slowly after the closure, and stays almost as high as the Newtonian peak for a very long time, which may increase the damage on the pulmonary epithelial cells.
      \item Higher stiffness ($G$) of the COPD and CF mucus causes smaller curvature at the capillary wave, and consequently smaller peaks of local normal stress gradients during the closure.
      \item High EVP characteristics in the COPD and CF conditions results in a slower accumulation of liquid at the axial center of the domain. This decreases the local tangential shear stress peaks during the closure.
    \end{itemize}
\end{comment}

These remarks suggest that the EVP model used in this paper can capture additional physics that have not been reported before. The increased peak of the stresses due to the extra stress contribution and the persisting extra stress after the closure may be especially important from the medical point of view, since they induce significant stress levels for the airway epithelium. As it is already known that CF and COPD can alter the mucus and surfactant secretion routines in the lungs \citep{agudelo2020decreased, griese1997pulmonary}, it is likely to observe airway closure in these patients. To decrease the damaging effects discussed above, therapeutic approaches such as inhaled hypertonic saline that aim to decrease the elasticity of airway mucus can be applied \citep{fahy2010airway}. Moreover, further research can discuss liquid plug propagation and rupture in these conditions to analyse whether liquid plug formation damages the pulmonary epithelium repeatedly.

\begin{acknowledgments}
Support from the Scientific and Technical Research Council of Turkey (TUBITAK), grant number 119M513, National Institutes of Health (NIH), grant number HL136141, and Business Finland E3, grant 4917/31/2021 is kindly acknowledged.
\end{acknowledgments}

% If in two-column mode, this environment will change to single-column format so that long equations can be displayed. 
%\begin{widetext}
%$$\mbox{put long equation here}$$
%\end{widetext}

% Figures should be put into the text as floats. 
% Use the graphics or graphicx packages (distributed with LaTeX2e).
% Put the label that you will use with \ref{} command in the braces of the \label{} command.
%
% \begin{figure}
% \end{figure}

% \begin{table}
% \end{tabular}
% \end{table}

%\begin{acknowledgments}
% Put your acknowledgments here.
%\end{acknowledgments}

\bibliography{PoF}

\end{document}